\documentclass[reprint,twocolumn,english,prl,notitlepage,nofootinbib,floatfix,longbibliography,aps]{revtex4-2}
\usepackage{siunitx}
\usepackage{graphicx}
\usepackage{placeins}
\usepackage{textcomp}
\usepackage{amssymb}
\usepackage{amsmath}
\usepackage{comment}
\usepackage{makecell}
\usepackage{physics}
\usepackage[english]{babel}
\usepackage{xcolor}
\usepackage{tabularx} 
\usepackage{adjustbox}
\usepackage{float}
\usepackage{booktabs} 
\usepackage[colorlinks=true,linkcolor=blue,urlcolor=blue,citecolor=blue,pdfusetitle]{hyperref}
\usepackage{upgreek}
\usepackage{xr}

\usepackage{soul}

\usepackage{times, txfonts}
\newcommand{\RR}[2]{\parbox[t]{#1}{\raggedright #2\strut}}
\newcommand{\RL}[2]{\parbox[t]{#1}{\raggedleft  #2\strut}}

\begin{document}

\title{Entanglement-enhanced optical magnetometry beyond the standard quantum limit}
\author{Jun Jia}
\thanks{These authors contributed equally}
\affiliation{Niels Bohr Institute, University of Copenhagen, Blegdamsvej 17, DK-2100 Copenhagen Ø, Denmark}
\author{Túlio Brito Brasil}
\thanks{These authors contributed equally}
\affiliation{Niels Bohr Institute, University of Copenhagen, Blegdamsvej 17, DK-2100 Copenhagen Ø, Denmark}
\author{Maimouna Bocoum}
\altaffiliation{Present address: Institut Langevin, ESPCI Paris, Université PSL, Sorbonne Université, Université Paris Cité, CNRS, 75005 Paris, France.}
\affiliation{Niels Bohr Institute, University of Copenhagen, Blegdamsvej 17, DK-2100 Copenhagen Ø, Denmark}
\author{Andrea Grimaldi} 
\altaffiliation{Present address: INFN, Sezione di Padova, I-35131 Padova, Italy.}
\affiliation{Niels Bohr Institute, University of Copenhagen, Blegdamsvej 17, DK-2100 Copenhagen Ø, Denmark}
\author{Laurits Møberg} 
\affiliation{Niels Bohr Institute, University of Copenhagen, Blegdamsvej 17, DK-2100 Copenhagen Ø, Denmark}
\author{Mikhail Balabas}
\affiliation{Niels Bohr Institute, University of Copenhagen, Blegdamsvej 17, DK-2100 Copenhagen Ø, Denmark}
\author{J\"{o}rg Helge M\"{u}ller}
\affiliation{Niels Bohr Institute, University of Copenhagen, Blegdamsvej 17, DK-2100 Copenhagen Ø, Denmark}
\author{Emil Zeuthen}
\affiliation{Niels Bohr Institute, University of Copenhagen, Blegdamsvej 17, DK-2100 Copenhagen Ø, Denmark}
\author{Eugene Simon Polzik}
\affiliation{Niels Bohr Institute, University of Copenhagen, Blegdamsvej 17, DK-2100 Copenhagen Ø, Denmark}

\begin{abstract}
Optical atomic magnetometry is a powerful tool for continuous sensing applications, yet, in the absence of quantum correlations, its sensitivity is limited by the standard quantum limit (SQL) stemming from a trade-off between optical probe imprecision and quantum measurement backaction. Beyond-SQL sensitivity requires quantum correlations that modify these measurement noise sources.  
Here we demonstrate such sensitivity by using entangled state of the probe light and by engineering correlations between measurement imprecision and backaction. Having first explored SQL in a broad range of frequencies, we demonstrate overcoming the limit by combining variational readout with coupling the magnetometer to one mode of a bipartite entangled light state and conditioning the results on the other entangled mode.  Tuning the detected light quadratures and combining the signals from the two measurement channels, we achieve sensitivity beyond the SQL in a broad range of acoustic frequencies which has so far remained inaccessible to quantum-noise-limited optical magnetometry. 
\end{abstract}

\maketitle

\section{Introduction}
The central problem of continuous weak-field sensing in the quantum-limited regime is determining a signal in the presence of measurement-induced disturbance of the sensor. In standard linear measurements \cite{clerk2010introduction, braginsky1995quantum,markus}, quantum mechanics imposes a trade-off between measurement imprecision, arising from probe phase shot noise, and quantum backaction originating from probe amplitude fluctuations acting on the sensor.  Their reciprocal scaling with probe strength and distinct frequency dependences gives rise to the standard quantum limit (SQL) \cite{braginsky1967classical, khalili2021quantum}, which defines the minimum added measurement noise achievable with uncorrelated quantum noise. In the simple case of a harmonic oscillator, the SQL is reached when the probe strength balances the two noise contributions at a given signal sideband frequency, yielding the best continuous-measurement sensitivity \cite{markus, Danilishin2012}. 
In particular, this quantum limit arises in optical atomic magnetometry \cite{budker2007optical,Budker_Jackson_Kimball_2013, wasilewski2010quantum}, where the measurement imprecision associated with the probe's phase quadrature noise competes with quantum backaction arising from vector AC-Stark shifts induced by fluctuations of the probe's amplitude quadrature \cite{fleischhauer2000quantum,vasilyev2012quantum}. The probe's phase and amplitude quadratures are conjugate variables satisfying $\comm{X(\Omega)}{P(\Omega')} = i\delta(\Omega+\Omega')$, so their non-commutation enforces an uncertainty relation linking imprecision and backaction. Reducing quantum imprecision by increasing probe strength inevitably enhances backaction noise in the absence of correlations; for each component of the frequency-dependent sensor response, this leads to an optimal trade-off between peak sensitivity and measurement bandwidth at a particular probe strength. Even with the probe strength optimized at each frequency, the added measurement noise remains bounded by the SQL, which constrains both the maximum achievable sensitivity and detection bandwidth.
\begin{figure*}[htp]
\centering
\includegraphics[width=0.9\textwidth]{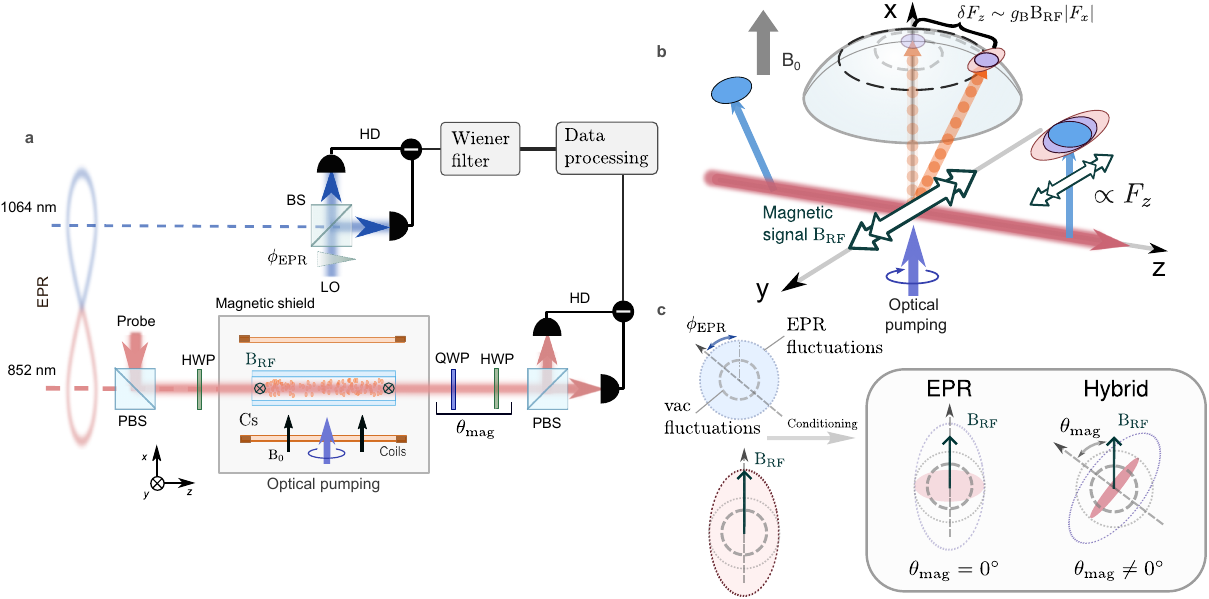}
\caption{\textbf{Principles of the entanglement-enhanced optical magnetometer and quantum-noise cancellation.} \textbf{a,} The input state of the pair of light beams (1064 and 852\,nm) is EPR-entangled. The probe beam at 852 nm interacts off-resonantly with an ensemble of Cs atoms optically pumped parallel to the dc magnetic field $B_0$. Quantum noise in the quadrature $\theta_{\mathrm{mag}}$, defined by the quarter waveplate (QWP), the half-wave plate (HWP), and polarizing beam splitter (PBS), is detected by a polarization homodyne detector (HD). The quadrature of the entangled reference beam at $1064$\,nm set by the phase shift $\phi_{\mathrm{EPR}}$ and measured by a second HD. A Wiener filter optimally suppresses quantum measurement noise to enhance magnetic field sensitivity. \textbf{b,} The spin ensemble polarized along the $x$-axis detecting radio-frequency magnetic fields ($\mathrm{B}_{\mathrm{RF}}$) applied along the $y$-axis. These RF signals tilt the spins along the $z$-axis, while the dc field leads them to precess around the $x$-axis. A linearly polarized probe laser measures the precessing spin projection onto the $z$-axis, including the intrinsic quantum uncertainty of the spins, via the Faraday effect. The measurement process introduces quantum noise from optical vacuum fluctuations. These fluctuations cause readout imprecision (shot noise) and disturb the atomic spins (quantum backaction). The transmitted light is then measured by a balanced polarimeter shown in a. \textbf{c,} Phasor diagrams of the light quantum noise before and after conditioning. Dashed grey circles mark the vacuum level, filled areas indicate the measured fluctuations, and the bold arrow indicates the displacement due to the magnetic signal $\mathrm{B}_{\mathrm{RF}}$. Left: noise of the two entangled arms measured individually — the 1064-nm reference beam in the quadrature set by $\phi_{\mathrm{EPR}}$ (blue) and the 852-nm probe after the atomic interaction (pink). Right (boxed): the conditioned noise obtained by combining the two channels. For $\theta_{\mathrm{mag}}=0^\circ$ (EPR) the noise is reduced along the signal direction; for $\theta_{\mathrm{mag}}\neq0^\circ$ (hybrid) the variational readout rotates the measured quadrature, and the combination suppresses the noise further.}
\label{fig:setup}
\end{figure*}

While the SQL marks the transition to quantum-limited measurement, where classical noise and the sensor's thermal noise become negligible, it does not represent a fundamental bound on linear measurement sensitivity. One approach to surpassing the SQL is quantum nondemolition (QND) measurement, in which the measured observable commutes with the system Hamiltonian \cite{Braginskii1996}. Examples include measurement of momentum-like quadratures \cite{braginsky1995quantum}, stroboscopic probing of atomic spins \cite{vasilakis2015generation}, and Bell-Bloom configurations that realize commuting spin observables \cite{colangelo2017simultaneous,troullinou2021squeezed}. Alternatively, commuting Einstein-Podolsky-Rosen (EPR) observables can be engineered via coupling to an effective negative-frequency reference frame, enabling joint measurement of both quadratures \cite{Polzik2015,Muller2017}.  Beyond commuting-observable strategies, sub-SQL sensitivity can be achieved by utilizing quantum correlations between imprecision and backaction. Such correlations allow quantum noise reduction, and have enabled sub-SQL displacement sensitivity in a mechanical membrane resonator \cite{kampel2017improving,mason2019continuous}, squeezed-light generation \cite{baerentsen2024squeezed}, and broadband quantum-noise suppression in gravitational-wave detectors through frequency-dependent squeezed-vacuum injection \cite{yap2020broadband,Ganapathy2023, jia2024squeezing}.

Here we demonstrate a room-temperature optical macroscopic atomic magnetometer achieving sensitivity beyond the SQL through engineered quantum cross-correlations. The idea and main components of the experiment are shown in Fig.~\ref{fig:setup}a. By combining variational readout with EPR-based conditioning of an entangled probe, we obtain sub-SQL 
sensitivity around selected frequencies, reaching noise levels below the minimum achievable with uncorrelated quantum noise. Importantly, the hybrid scheme enables tunable control of the depth, central frequency, and bandwidth of the sensitivity enhancement mitigating the conventional sensitivity–bandwidth trade-off \cite{Danilishin2019,clerk2010introduction}. 
Extending quantum-enhanced magnetometry into the low-acoustic regime—relevant for biomagnetic \cite{Aslam2023,hamalainen1993magnetoencephalography} and geomagnetic detection \cite{stuart1972earth,rikitake1968geomagnetism,glenn2017micrometer}—has remained challenging due to dominant technical noise, and most previous demonstrations of sub-SQL magnetometry have been restricted to operation at higher Larmor frequencies in the radio-frequency band \cite{wasilewski2010quantum}. Here we overcome this limitation by demonstrating $2.1\pm 0.3$\,dB sub-SQL performance down to the low-acoustic regime ($\sim$ 7\,kHz) in a room-temperature atomic platform. These results bridge the long-standing gap between quantum-enhanced measurement and application-relevant low-frequency magnetometry.

\section{Magnetometer and its standard quantum limit of sensitivity}
 We implement the orientation-based, room-temperature optical magnetometer illustrated in Fig.~\ref{fig:setup}b.  An ensemble of optically polarized atoms is subjected to a DC bias magnetic field along the $x$-axis, inducing spin precession about this quantization axis at the Larmor frequency $\Omega_\mathrm{S}$.  
External magnetic fields oscillating at frequencies $\Omega_{k}$, near $\Omega_{S}$, drive the transverse spin y-component.  The resulting spin precessions imprint polarization modulations onto the linearly polarized probe light via the Faraday interaction acting as a highly sensitive transducer for the external magnetic fields perturbations \cite{hammerer2010quantum, geremia2006tensor}. The probe light after interaction is detected via a balanced polarimeter (Fig.~\ref{fig:setup}a).
\begin{figure*}[htp]
\centering
\includegraphics[width=1\textwidth]{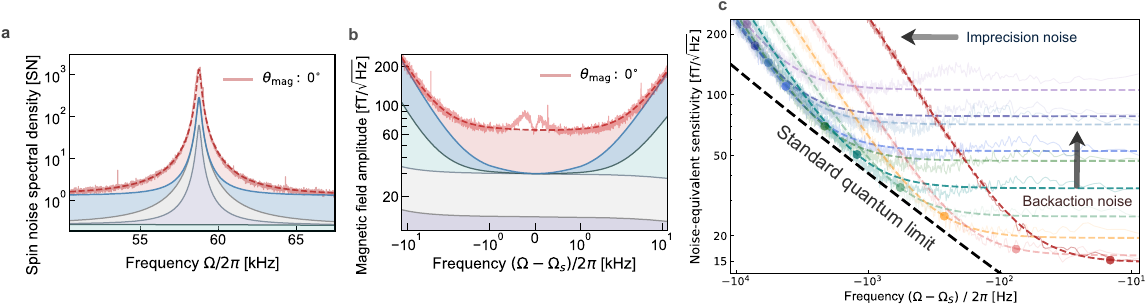}
\caption{\textbf{ Noise budget and sensitivity limits of an RF orientation-based optical magnetometer.} \textbf{a,} Measured phase quadrature ($\theta_{\text{mag}}$ = 0$^{\circ}$) of the spin noise spectrum for a vacuum-noise-limited probe (red trace) in probe shot noise units. The red dashed curve shows the fitted spin noise and the shaded areas present individual noise contributions: quantum backaction (red), measurement imprecision (blue), spin thermal noise (grey), spin projection (purple), and spin broadband (teal). \textbf{b,} Magnetic field sensitivity reconstructed using quantum noise data from \textbf{a}. Coloured areas with the same colour code as in \textbf{a} highlight the dominant noise contributions across the entire frequency range. \textbf{c,} Probe-power dependence of the reconstructed magnetic-field sensitivity. Colored circles indicate the frequency ranges where the sensitivity at that specific probe power outperforms the other configurations. The predicted SQL (dashed black line) is calculated from independently calibrated system parameters using the dark decoherence rate and transduction factor calibrated at 1\,mW probe.}
\label{fig:spin setup and noise spectrum}
\end{figure*}

The magnetic-field sensitivity is determined by the total magnetometer noise referred to the input magnetic field. In the linear-response regime, the noise equivalent magnetic-field spectral density can be written as
\begin{equation}\label{eqM:variational_readout}
S_{\mathrm{B}}(\Omega)= \frac{
S_{\mathrm{imp}}(\Omega)+S_{\mathrm{qba}}(\Omega)+S_{\mathrm{corr}}(\Omega)+S_{\mathrm{spin}}(\Omega)}{A_{\mathrm{B}}\Gamma_\mathrm{S} \abs{\rho_{\mathrm{S}}(\Omega)}^{2}\cos^2(\theta_{\mathrm{mag}})},
\end{equation}
where $S_{\mathrm{imp}}$, $S_{\mathrm{qba}}$, and $S_{\mathrm{corr}}$ denote measurement imprecision, quantum backaction, and their correlation, respectively, and $S_{\mathrm{spin}}$ represents the intrinsic spin fluctuations (see details in Methods).
 The experimentally calibrated magnetometer transfer function $\sqrt{A_{\mathrm{B}}\Gamma_{\mathrm{S}}} \rho_{\mathrm{S}}(\Omega)$ describes the conversion of magnetic-field amplitude into collective spin displacement with the transduction factor $\sqrt{A_{\mathrm{B}}}$, and the dynamical spin response governed by the susceptibility $\rho_\mathrm{S}(\Omega)$, followed by optical readout of the spin quadrature $\theta_{\mathrm{mag}}$ set by the waveplates (Fig.~\ref{fig:setup}a) at a rate $\Gamma_{\mathrm{S}}$ (see Methods). The noise-equivalent magnetic-field sensitivity is given by $\delta B(\Omega) = \sqrt{S_{\mathrm{B}}(\Omega)}$. In the absence of correlation ($S_{\mathrm{corr}} = 0$), realized for $\theta_{\mathrm{mag}} = 0^{\circ}$, minimizing the probe-related added noise defines the standard quantum limit (SQL): $S_{\mathrm{B}}(\Omega)\geq S_{\mathrm{B, SQL}}(\Omega) = \abs{\chi_{\mathrm{S}}(\Omega)}/(A_{\mathrm{B}}\abs{\rho_{\mathrm{S}}(\Omega)}^{2})$, which depends only on the atomic susceptibility functions $\chi_\mathrm{S} (\Omega)$, $\rho_\mathrm{S} (\Omega)$ and the field-to-displacement transduction factor $A_{\mathrm{B}}$ set by the collective spin length. Surpassing the SQL requires engineering correlations between imprecision and backaction that modify the effective measurement noise. In this work, we explore three strategies for achieving this: variational readout exploits intra-channel correlations; EPR conditioning uses inter-channel correlations with an entangled reference field; and a hybrid scheme combines both mechanisms.

\section{Noise budget and reference measurements with near-SQL sensitivity}
 The magnetometer quantum limited sensitivity and noise contributions are quantified through the power spectrum density (PSD) as depicted in Fig.~\ref{fig:spin setup and noise spectrum}a, measured without the EPR-entangled source injection. To characterize the frequency-dependent magnetometer response, we first set the Larmor frequency to $\Omega_{\text{S}}/2\pi\approx 59$\,kHz and apply multi-tone RF signals of equal amplitude, enabling simultaneous measurement across multiple frequencies.
As detailed in Methods, the calibrated spin oscillator model decomposes the total noise spectrum into intrinsic spin fluctuations (thermal and projection noise) and measurement-associated noise due to quantum imprecision and backaction (Fig.~\ref{fig:spin setup and noise spectrum}a). The magnetic peak sensitivity  64\,$\text{fT}/\sqrt{\text{Hz}}$ limited by these noise sources in a 3\,dB detection bandwidth of $\approx$ 8\,kHz is shown in Fig.~\ref{fig:spin setup and noise spectrum}b. The uncorrelated spin noise measured at the phase quadrature ($\theta_\mathrm{mag} = 0^\circ$) defines the reference baseline (labelled ``Uncorr.\ ref.'').  Near the Larmor frequency, quantum backaction noise (red) limits the magnetometer peak sensitivity, whereas imprecision noise (blue) governs the off-resonant response and therefore constrains the detection bandwidth. Beyond imprecision, the detection bandwidth is further limited by broadband spin-noise components (teal), which arise from the faster decay of additional spin modes during readout \cite{borregaard2016scalable,Shaham2020}. The intrinsic spin thermal (gray)/projection (purple) noise sets the next boundary for further sensing improvement.

Fig.~\ref{fig:spin setup and noise spectrum}c benchmarks the reconstructed noise-equivalent field sensitivity against the SQL across varying probe strength. At a weak probe power of 10\,$\upmu W$, the platform reaches a peak sensitivity of approximately 15\,$\text{fT}/\sqrt{\text{Hz}}$  over a  $\approx$ 120\,Hz bandwidth. Increasing the probe power reduces the imprecision contribution and extends the detection bandwidth but elevates quantum backaction, illustrating the SQL trade-off. The envelope of these power-optimized minima approaches the predicted SQL in the range of detuning  $\approx$ (1--5) kHz from resonance,  confirming near-quantum-limited operation. However, further increasing the probe strength causes the sensitivity to progressively departs from the SQL, as the collective spin length, and hence the transduction factor $A_\mathrm{B}$, degrades (see Extended Data Fig.~\ref{fig:SQL validation} in Methods). This bottleneck motivates enhancing the sensitivity via quantum correlations rather than probe power scaling, as explored below. 
\section{Sensitivity enhancement via variational readout}
\begin{figure}[htp]
    {\centering    \includegraphics[width=0.48\textwidth]{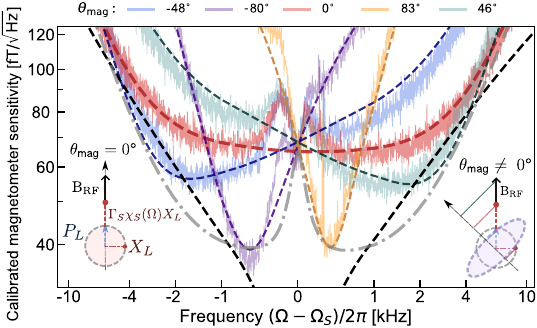}
    \caption{\textbf{Quantum-enhanced optical magnetometer enabled by variational readout.} The output-light phasor diagrams illustrate the quantum enhancement mechanism of the variational readout. At $\theta_{\mathrm{mag}} = 0^{\circ}$ (left inset), imprecision $(P_{L})$ and backaction $(\Gamma_\mathrm{S}\chi_\mathrm{S}(\Omega)X_{L})$ are uncorrelated, add incoherently, and lie along the signal ($B_{\mathrm{RF}}$) limiting the sensitivity. Rotating the detection quadrature away from the pure phase quadrature ($\theta_{\mathrm{mag}} \neq 0^{\circ}$, right inset) utilizes the backaction-imprecision correlation -- the ponderomotive squeezing (purple ellipse) -- to reduce the measurement noise. Sensitivity enhancement occurs when the noise reduction outweighs the accompanying ${\cos}(\theta_{\mathrm{mag}})$ signal loss. The colour traces represent the total-noise equivalent magnetic sensitivity at different detection quadratures. Dashed curves are theoretical predictions based on experimentally calibrated parameters. The grey dash-dotted curve indicates the frequency-optimized sensitivity while the black dashed lines represent the inferred standard quantum limit (SQL). We define the 3\,dB detection bandwidth relative to the minimum sensitivity achieved for a given fixed detection phase. The quantum-enhancement window is the frequency range where the sensitivity surpasses the uncorrelated readout reference ($\theta_{\text{mag}}= 0^{\circ}$). The Larmor frequency $\Omega_\mathrm{S}/2\pi$ is set to 58.8\,kHz. }
    \label{fig:quantum enhancement with variational readout}}
\end{figure}

With the noise sources calibrated, we proceed with the first approach to quantum-enhanced sensitivity by using variational readout and virtual rigidity. The dashed black straight lines in Fig.~\ref{fig:quantum enhancement with variational readout} indicate the SQL, corresponding to the minimal added uncorrelated quantum measurement noise achieved with the optimized probe strength at each frequency (see Methods for detailed SQL calibration).
Rotating the atomic detection away from the pure phase quadrature ($\theta_{\mathrm{mag}} \neq 0 ^{\circ}$) with the help of the waveplates shown in Fig.~\ref{fig:setup}a introduces non-zero imprecision-backaction correlations, modifying both the magnetic signal amplitude and the total added quantum noise \cite{baerentsen2024squeezed}, an effect known as ponderomotive squeezing in opto-mechanics \cite{kampel2017improving,mason2019continuous}. The impact on the sensitivity $\delta B(\Omega)$ is largely captured by the concept of virtual rigidity (see, e.g., Ref.~\cite{Zeuthen2019}): Part of the (nominal) readout rate is sacrificed, leading to effective rate $\Gamma_\mathrm{S}\rightarrow\cos^2(\theta_\mathrm{mag})\Gamma_\mathrm{S}$, in order to achieve an effective frequency shift of the magnetometer resonance $\Omega_\mathrm{S}$ by $\approx{\sin} (2\theta_\mathrm{mag})\Gamma_\mathrm{S}/4$ [valid for $\Omega^2\approx\Omega_\mathrm{S}^2\gg(\gamma_\mathrm{S}/2)^2$]. Hence, changing $\theta_\mathrm{mag}$ can serve as a flexible means of effectively shifting $\Omega_\mathrm{S}$ without having to adjust the DC magnetic bias field, at the expense, however, of exacerbating the impact of detection losses (see Supplementary Material for details).
At $\Omega/2\pi\approx58$\,kHz (Fig.~\ref{fig:quantum enhancement with variational readout}), approximately 5\,dB of ponderomotive squeezing enhances the magnetic field sensitivity from 64 to 38\,$\text{fT}/\sqrt{\text{Hz}}$ (purple curve), corresponding to a factor of $\approx 1.7$ improvement relative to the uncorrelated readout reference (red curve). The enhancement occurs within a $\approx$ 1.2\,kHz window -- defined as the frequency range where the sensitivity surpasses this reference -- slightly detuned from the Larmor frequency, with the expected symmetric behaviour above resonance when the detection phase is inverted (orange trace). This quantum enhancement window can be extended to approximately 6\,kHz by tuning the detection phase (blue/teal traces).
For a fixed detection phase, however, improving peak sensitivity generally narrows the 3\,dB detection bandwidth of that operating point, reflecting the underlying sensitivity-bandwidth trade-off. Allowing the detection phase to vary with frequency instead tracks the optimal correlation angle across the spectrum. The dash-dotted gray curve projects the frequency-optimized sensitivity based on calibrated parameters, providing a path to mitigate this trade-off and approach sub-SQL performance over a broader band. 

 \begin{figure*}[htp]
\centering
\includegraphics[width=1\textwidth]{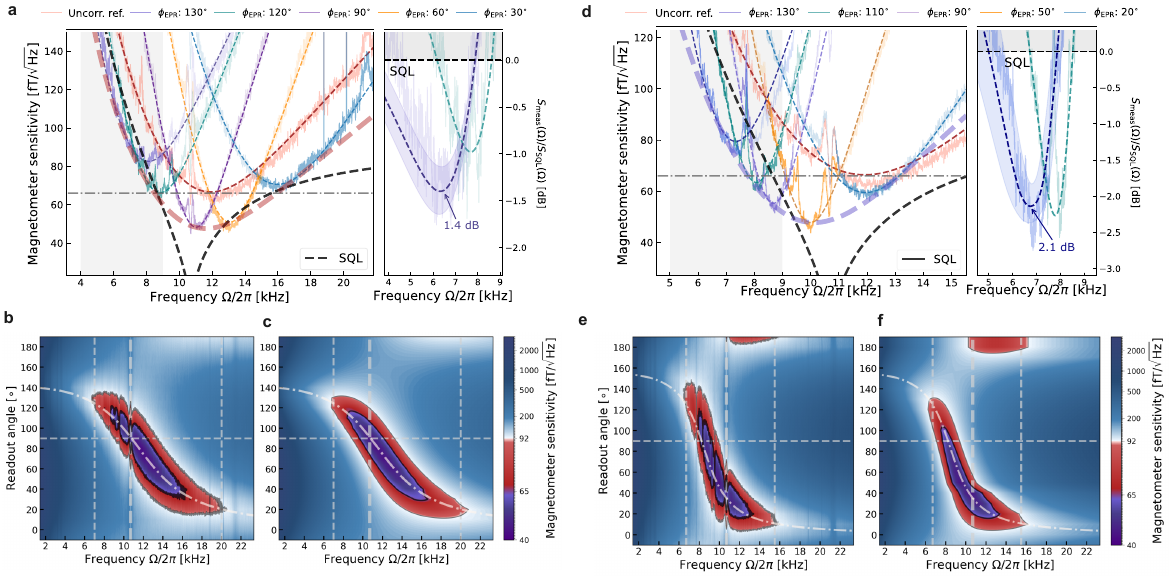}
\caption{\textbf{Entanglement-enhanced optical magnetometer.} \textbf{a,} Entanglement injection and no variational readout, the magnetometer detection phase phase ($\theta_{\text{mag}} = 0^{\circ}$)  while the 1064-nm reference mode is measured with a tunable phase ($\phi_{\text{EPR}}$). The plot shows the noise-equivalent magnetic-field sensitivity spectra calibrated for a Larmor precession of 10.7\,kHz. The uncorrelated spin noise reference (light red trace) yields a peak sensitivity of 65\,$\text{fT}$/$\sqrt{\text{Hz}}$ (dash-dotted grey line). Adjusting $\phi_{\text{EPR}}$ (colored traces) allows for frequency-tunable quantum noise reduction, well-matched by theoretical predictions (dashed curves). The thick red dashed curve denotes the combination of the frequency-optimized sensitivity. The right inset details the noise normalized to the standard quantum limit (SQL), demonstrating a maximum sub-SQL sensitivity enhancement of 1.4 $\pm$ 0.2\,dB at $\phi_{\text{EPR}}=130^{\circ}$.  \textbf{b \& c,} Experimental (b) and theoretical (c) contour maps detailing the noise-equivalent sensitivity across the parameter space of sideband frequency and EPR homodyne phase. The shaded regions highlight regimes where the sensitivity surpasses the uncorrelated readout reference (purple). The red regions mark frequencies where this quantum enhancement extends beyond the reference's conventional 3\,dB bandwidth (defined by a sensitivity threshold of 92\,$\text{fT}/\sqrt{\text{Hz}}$). The white dash-dotted trace indicates the optimal frequency-dependent conditioning phase required to achieve the broadband enhancement shown in (a). \textbf{d,} Entanglement injection and magnetometer detection phase tuned to ($\theta_{\text{mag}} \approx -55^\circ$) to enable variational readout. Noise-equivalent magnetic-field sensitivity spectra as a function of the EPR-conditioning homodyne phase ($\phi_{\text{EPR}}$). The uncorrelated spin noise reference (light red trace) is compared against the dynamically frequency-optimized sensitivity (thick purple dashed curve), which shifts the enhancement window further into the acoustic regime. The right inset shows the conditional variance normalized to the SQL, demonstrating a maximum noise suppression of $2.1 \pm 0.3$\,dB below the SQL at $\phi_{\text{EPR}} =130^\circ$.  \textbf{e \& f} Experimental (e) and theoretical (f) contour maps detailing the noise-equivalent field sensitivity across sideband frequencies and EPR readout angles. The purple region denotes sensitivities surpassing the uncorrelated vacuum reference, with the optimal performance shifted 2\,kHz below the reference's natural peak. The red region highlights the expansion of the 3\,dB sensitivity bandwidth beyond that of the standard reference magnetometer.}
\label{fig:spin_noise_spectrum_with_EPR_conditioning}
\end{figure*}

The SQL and frequency-optimized sensitivity curves are reconstructed using the transduction factor averaged over five independently calibrated detection angles, whose residual variation is attributed primarily to atom-number fluctuations during the measurements (see Methods). At resonance, the real part of backaction-imprecision cross-correlation vanishes, so variational polarimeter homodyne detection cannot reduce quantum noise there. In principle, on-resonance sub-SQL performance can instead be achieved through access to the complex cross-correlation provided by synodyne detection, at the price of retrieving only one phase component of the signal \cite{buchmann2016complex,kampel2017improving}.

\section{Entanglement-enhanced sensitivity at acoustic frequencies}
Having demonstrated sensitivity enhancement through variational readout, we proceed to using an entangled state of the probe light in addition to engineered correlations between measurement imprecision and backaction.
Motivated by the relevance of magnetometry in the kilohertz frequency range for biomedical and other applications, we tune the atomic Larmor frequency down to 10\,kHz. Minimizing the probe classical noise using an active intensity noise eater \cite{Alkis2023thesis} enables the observation of approximately 5\,dB ponderomotive squeezing in this range with a vacuum-noise-limited probe.
We then replace the vacuum mode in the orthogonal probe polarization with one arm of an Einstein-Podolsky-Rosen (EPR) entangled optical source ($\approx 7$\,dB two-mode squeezing) (see Extended Data Fig.~\ref{fig:extend_figure_4}) and perform conditional measurements using the correlated parallel channel (Fig.~\ref{fig:setup}a). EPR conditioning reduces the added measurement noise (Fig.~\ref{fig:setup}b) through cross-correlations with the optical reference channel by preparing a (conditional) single-mode squeezed state in the magnetometer channel (Fig.~\ref{fig:setup}c). These correlations are currently narrowband due to the fixed conditioning quadrature matched to the spin response function for particular signal frequencies $\Omega$. Frequency-dependent conditioning would allow spin-light correlations to be harnessed across a broad acoustic band. Since the conditioning does not alter the spin response or decoherence rate, the conditional noise-equivalent sensitivity spectra can be reconstructed using the previously calibrated transfer function. 

Figure~\ref{fig:spin_noise_spectrum_with_EPR_conditioning}a shows the reconstructed noise-equivalent sensitivity spectra for varied EPR homodyne phase $\phi_\mathrm{EPR}$, which determines the angle of the conditionally squeezed quadrature, with the atomic phase quadrature fixed at $\theta_{\mathrm{mag}} = 0^{\circ}$. EPR conditioning yields up to 3\,dB noise suppression relative to the uncorrelated readout reference noise labelled as ``Uncorr.\ ref.'' in Fig.~\ref{fig:spin_noise_spectrum_with_EPR_conditioning}a. Far from resonance, where imprecision dominates, the correlated EPR phase ($\phi_{\mathrm{EPR}}= 0^{\circ}$) aligns with the atomic channel phase quadrature. Closer to the Larmor frequency, where quantum backaction dominates, the optimal phase shifts toward the orthogonal quadrature ($\phi_{\mathrm{EPR}}= 90^{\circ}$). This phase rotation reflects the changing balance between imprecision and backaction across the sideband frequency range. It improves the peak sensitivity from 65 to approximately 47\,$\mathrm{fT}/\sqrt{\mathrm{Hz}}$. The achieved sensitivity at around $\phi_{\mathrm{EPR}} = 130^{\circ}$ lies below the SQL (black dashed curve), confirming $1.4\pm 0.2$\,dB sub-SQL performance enabled by EPR correlations.

With a fixed conditioning phase $\phi_\mathrm{EPR}$, however, the squeezing angle does not vary with Fourier frequency $\Omega$; so the quantum enhanced sensitivity with EPR conditioning is confined to a limited spectral window. Allowing the conditioning phase to vary with $\Omega$ (following the white dash-dotted curve in Fig.~\ref{fig:spin_noise_spectrum_with_EPR_conditioning}b\&c), as could be realized by implementing an $\Omega$-dependent quadrature rotation in the reference channel, extends the enhancement across an $\approx$ 8\,kHz window where the sensitivity surpasses the uncorrelated readout reference (purple region) and broadens the performance beyond the conventional 3\,dB bandwidth of that reference configuration from $\approx$ 8\,kHz to $\approx$ 13\,kHz (red region). Below 5\,kHz, residual classical noise limits further improvement, as indicated by the discrepancy between experiment and theory, although quantum enhancement remains observable down to 3\,kHz (see Extended Data Fig.~\ref{fig:extend_figure_4}).
Without the more complex detection scheme required for variational readout, frequency-optimized EPR conditioning provides a direct framework for broadband quantum-enhanced sensing.

\begin{figure}[htbp]
\centering
\includegraphics[width=0.48\textwidth]{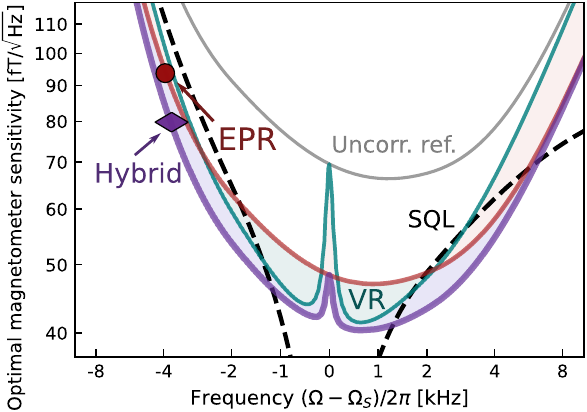}
\caption{\textbf{Optical magnetometer with frequency-optimized detection and EPR conditioning phases.} Teal, red, and purple curves show the predicted quantum-enhanced magnetometer sensitivity for variational readout (VR), EPR conditioning, and hybrid configuration, respectively, based on the experimentally calibrated parameters. The shaded regions indicate the frequency ranges in which each configuration outperforms the others. Diamond and circle markers denote the experimentally observed sensitivity enhancement discussed in this work. The grey curve is the uncorrelated, vacuum limited spin noise reference, while the dashed black line indicates the SQL. The Larmor frequency $\Omega_\mathrm{S}/2\pi$ is fixed at 10.7\,kHz. The figure highlights the complementary roles of variational readout and EPR conditioning, and their joint enhancement of sensitivity in the hybrid scheme.}
\label{fig:optimal sensing}
\end{figure}

The two complementary approaches, the variational readout and the EPR enhancement, can be combined to further engineer the quantum measurement noise. In the hybrid configuration (Fig.~\ref{fig:spin_noise_spectrum_with_EPR_conditioning}d), the atomic channel is detected at an intermediate quadrature ($\theta_{\mathrm{mag}} \approx -55^{\circ}$, extracted from fit), exploiting the ponderomotive squeezing of the EPR fluctuations. The residual quantum noise is further suppressed through the EPR-based conditional inference and the transfer function is recalibrated for this intermediate detection quadrature to account for the corresponding reduction in signal gain.
The reconstructed sensitivity spectrum Fig.~\ref{fig:spin_noise_spectrum_with_EPR_conditioning}d shows a peak sensitivity of 47\,$\mathrm{fT}/\sqrt{\mathrm{Hz}}$, corresponding to $\approx$5\,dB quantum noise reduction at $\sim$2\,kHz below the peak-sensitivity frequency of the uncorrelated readout. Around 7\,kHz, the hybrid scheme exhibits $2.1\pm 0.3$\,dB noise suppression below the SQL. This hybrid configuration yields an improved minimum sensitivity compared to variational readout at the same angle. It also shifts the optimal sensitivity toward lower frequencies compared with EPR conditioning.
Frequency-optimized conditioning (thick dashed purple curve in Fig.~\ref{fig:spin_noise_spectrum_with_EPR_conditioning}d) confirms this shift, although the enhanced frequency window is $\approx3$\,kHz narrower (indicated by the purple region in Fig.~\ref{fig:spin_noise_spectrum_with_EPR_conditioning}e\&f) than that obtained with EPR conditioning alone. These results demonstrate that hybrid quantum correlation engineering enables controllable reshaping of the frequency-dependent sensitivity of the optical magnetometer.
Using the experimentally calibrated parameters and theoretical model, we reconstruct the optimal magnetometer sensitivity at each analysis frequency by numerically optimizing both the detection and conditioning phases. The resulting prediction (Fig.~\ref{fig:optimal sensing}) shows that the hybrid configuration combines the quantum enhanced regime of variational readout and EPR-conditioning, yielding improved signal sensitivity over a broad frequency range. The experimentally implemented operating point presented here corresponds to an intermediate variational readout angle with optimized EPR-conditioning, providing a direct demonstration of this combined enhancement, which confirms that joint optimization of intra and inter channel quantum correlations enables improvement of magnetometer sensitivity beyond each scheme alone.
\section{Conclusion and Outlook}
In this work, we demonstrate a sub-SQL magnetometer in a room-temperature atomic ensemble through engineered quantum correlations. By combining variational readout with entanglement-based conditional inference, we achieve control over both the depth and the spectrum of the sensitivity. This enables the optimal sensing frequency to be repositioned away from the intrinsic Larmor resonance rather than being fixed by the spin susceptibility. As seen in Fig.~\ref{fig:spin setup and noise spectrum}c, conventional power scaling cannot reach the SQL across the full band, being limited by quantum backaction and ultimately by the degradation of the collective spin length. The quantum-correlation engineering demonstrated here provides an alternative route, redistributing quantum measurement noise to achieve enhanced sensitivity over a several-kilohertz band and thereby relaxing the conventional trade-off between peak sensitivity and bandwidth. Such frequency-dependent squeezing could be realized, for example, by coupling the EPR source to a detuned filter cavity \cite{jia2024squeezing} or to an effective negative-mass reference oscillator \cite{novikov2025hybrid}. 

Radio-frequency quantum-enhanced macroscopic optical magnetometers typically operate at Larmor frequencies in the hundreds-of-kilohertz to megahertz range, whereas spin-exchange-relaxation-free (SERF) magnetometers achieve high sensitivity at ultra-low frequencies ($\leq$ 200\,Hz) but outside the quantum-limited measurement regime \cite{li2018serf}. Our results demonstrate quantum-limited operation at acoustic Larmor frequencies, extending sub-SQL magnetometry into this new domain. Further reduction of residual low-frequency technical and DC alignment noise, together with the implementation of frequency-dependent squeezing injection following the measured optimal-angle trend, could extend the demonstrated narrowband enhancement toward broadband operation compatible with biomagnetic signal bandwidths and sensitivity levels \cite{jin2026four,9102993}.

From a theoretical perspective, the achievable field sensitivity is weighted by the spin response function. Lowering the Larmor frequency into the acoustic regime enhances the anisotropy of the transverse spin transfer functions (see Supplementary Material). In particular, magnetic fields along the $y$ and $z$ directions couple through distinct susceptibilities, resulting in different noise-equivalent sensitivities away from resonance as shown in Extended Data Fig.~\ref{fig:Extended_Figure_5}. This implies that the quantum limited sensitivity itself becomes direction- and frequency-dependent, set by the underlying spin dynamics. The model therefore predicts enhanced sensitivity for alternative field components in specific frequency ranges. This anisotropy indicates that optimal broadband operation may be achieved either by multi-axis probing or by using a frequency-dependent measurement basis. More generally, these results show that engineered quantum correlations enable tunable, frequency-dependent sensitivity in optically probed macroscopic oscillators.

\newpage
\FloatBarrier
\newpage
\section{Methods}
\section{Experimental platform and operating conditions}
The experimental platforms, the atomic spin oscillator and the EPR-entangled light source, are described in detail in Refs.~\cite{Brasil2022, Jia2023, novikov2025hybrid}.
\textbf{Atomic spin oscillator:} The orientation-based optical magnetometer employs $\approx10^{10}$ $^{133}$Cs atoms confined in a $2\times2\times80$\,mm$^{3}$ paraffin-coated glass cell with anti-reflection-coated windows for 852-nm light. The atomic density is set by maintaining the cell temperature at $\approx40^{\circ}$C. The cell is enclosed in a five-layer magnetic shield to suppress the environmental magnetic noise. A homogeneous, DC bias magnetic field defines the spin Larmor frequency, while calibrated radio-frequency (RF) fields generated by a PCB coil are used to characterize the magnetometer frequency response. 
The atoms are optically pumped into the $|F=4, m_{F} = -4\rangle$ ground state with $\approx82 \%$ spin polarization using continuous repumping light. Spin precession driven by weak RF magnetic fields is detected via Faraday rotation of a probe beam blue-detuned by 1.6\,GHz from the D2-line transition. The probe power is fixed at 1\,mW, which defines the readout rate $\Gamma_{S}$ used throughout the measurements. The probe beam is shaped into a square top-hat profile with $80\%$ cell volume filling to ensure uniform light-atom coupling strength. Classical probe amplitude noise in the acoustic band is suppressed using an active noise eater~\cite{Alkis2023thesis}.
\textbf{EPR-entangled light source:} Nondegenerate EPR-entangled optical modes at 852\,nm and 1064\,nm are produced via parametric down-conversion of a 473\,nm pump in a bow-tie nondegenerate optical parametric oscillator containing a type-0 PPKTP crystal. Two entangled modes are separated by a dichroic mirror. The 852\,nm mode is combined with the probe light through a polarized beam splitter and detected using a balanced polarimeter, while the correlated 1064\,nm mode is measured by balanced homodyne detection with a coherent local oscillator. Conditional inference using the 1064\,nm channel yields approximately 4\,dB noise reduction relative to the atomic-arm imprecision noise down to 3\,kHz (Extended Data Fig.~\ref{fig:extend_figure_4}), corresponding to the squeezing factor $r\approx1.43$. The electronic noise of both detectors is more than 17\,dB below the shot-noise level. The degradation of conditional squeezing below 3\,kHz is mainly due to the quadrature-locking phase noise \cite{grimaldi2025coherent}, atomic alignment noise near DC \cite{Jia2023}, and residual probe amplitude noise coupled through the imperfect linear polarization of the probe \cite{jiathesis}.
\section{Linear-response and spectral-noise model of optical magnetometer}
With the engineered interaction via the probe input polarization angle \cite{geremia2006tensor, Jia2023} and linearization around a strongly polarized spin state, the system is described by the effective interaction Hamiltonian (in units where $\hbar=1$),
\begin{equation}
H_{\mathrm{eff}} = \sqrt{\Gamma_{S}}X_{S}X_{L}-\sqrt{A_{\mathrm{B}}}B_{\mathrm{RF}}P_{S},
\end{equation}
where $\Gamma_{S}$ is the probe readout rate. $A_{\mathrm{B}}$ is the probe-power-independent transduction factor mapping the applied RF-magnetic field onto the spin displacement, determined by the gyromagnetic ratio and the steady state of polarized collective spin length $\abs{F_{x}}=F^{ss}_{x}$; the operators $[X_{S},P_{S}]=i$ and $[X_{L}(t),P_{L}(t')]=i\delta(t-t')$ are dimensionless quadratures of the collective spin and the probe field, respectively (see details in Supplementary Material). The first term describes the Faraday interaction between probe and spin oscillator, while the second term represents the driving of the spin oscillator by the external RF magnetic field $B_{\mathrm{RF}}$ (assuming presently that it is linearly polarized along $y$). 
In the stationary, linear response regime, the spin dynamics are characterized by the two spin susceptibilities to influences acting on $X_S$ and $P_S$, respectively,
\begin{align}
\mathcal{\chi}_S(\Omega) &= \frac{\Omega_{S}}{\Omega^{2}_{S} + (\gamma_{S}/2)^2 -\Omega^{2} -i \gamma_{S}\Omega},\\
\mathcal{\rho}_S(\Omega) &= \frac{\gamma_{S}/2-i\Omega}{\Omega^{2}_{S} + (\gamma_{S}/2)^2 -\Omega^{2} -i \gamma _{S}\Omega},
\end{align}
governing, in particular, the response of the spin to probe backaction and the external magnetic field applied along the orthogonal transverse axis, respectively, including their decoherence-associated damping contributions. 
The detected probe light quadrature after interacting with atoms can be written as
\begin{equation}\label{eqM:scattering-rel}
\begin{aligned}
    P_{L}^{\mathrm{out}}(\Omega) &=P_{L}^{\mathrm{in}}(\Omega) + \Gamma_{S} \chi_{S}(\Omega)X_{L}^{\mathrm{in}}(\Omega) \\
    &\quad- \sqrt{\Gamma_{S}\gamma_{S}}(\rho_{S}(\Omega)f^{X_{S}}(\Omega)+\chi_{S}(\Omega)f^{P_{S}}(\Omega))\\
    &\quad+ \sqrt{\Gamma_{S}A_{\mathrm{B}}}\rho_{S}(\Omega)B_{\mathrm{RF}}(\Omega).
\end{aligned}
\end{equation}
The first line contains measurement imprecision (shot noise) and measurement-induced quantum backaction, the second line accounts for intrinsic spin fluctuations, and the last term describes the driven response to an external RF magnetic field (see Supplementary Information for details).
The corresponding measured noise spectrum is then
\begin{equation}
    S_{\mathrm{meas}}(\Omega)
    =
    S_{\mathrm{imp}}(\Omega)
    +
    S_{\mathrm{qba}}(\Omega)
    +
    S_{\mathrm{spin}}(\Omega)
    +
    S_{\mathrm{corr}}(\Omega),
\end{equation}
where the terms denote the respective contributions identified above, with the $S_{\mathrm{corr}}$ arising from correlations engineered by the measurement scheme.
The resulting noise-equivalent magnetic-field sensitivity is
\begin{equation}\label{eqM:delta-B_y}
    \delta B_{y}(\Omega)
    =
    \frac{\sqrt{S_{\mathrm{meas}}(\Omega)}}{\sqrt{\eta^{\mathrm{out}    }_{\mathrm{mag}}\Gamma_S A_{\mathrm{B}}}\,\abs{\rho_S(\Omega)}\cos(\theta_{\mathrm{mag}})}.
\end{equation}
Assuming the absence of correlations $S_\mathrm{corr}(\Omega)$ and spin noise $S_\mathrm{spin}(\Omega)$, minimizing the added quantum noise \eqref{eqM:delta-B_y} over $\Gamma_S$ for each Fourier component $\Omega$, separately, defines the standard quantum limit (SQL)
\begin{align}
    \delta B_{y,\text{SQL}}(\Omega)  = \sqrt{\abs{\chi_{S}(\Omega)}/(A_{B} \abs{\rho_{S}(\Omega)}^{2})}.
\end{align}
In the experiment, finite optical transmission and detection efficiency reduce observable quantum correlations and introduce additional vacuum noise, which are incorporated through an effective detection efficiency $\eta^{\mathrm{out}}_{\mathrm{mag}}$.
All spectra are expressed in canonical units where vacuum fluctuations correspond to $1/2$, while experimentally the data are normalized to the probe shot noise (unity). Accordingly, theoretical spectra are rescaled by a factor of 2 when fitting to the data.
The variational readout, EPR conditioning, and hybrid configurations differ primarily in the structure of the correlation term $S_{\mathrm{corr}}(\Omega)$, while the underlying spin response functions remain unchanged.
Explicit expressions for the transfer functions, loss mechanisms, probe power broadening, and configuration-dependent noise spectra used for fitting are provided in the Supplementary Material.

\section{Calibration of multi-tone-fields and magnetometer transfer function}
Following the realization of a quantum-noise-limited optical magnetometer, we calibrate its magnetic-field sensitivity by injecting controlled RF magnetic fields and analysing the atomic response in the power spectral density of the detected photocurrent.
\subsection{Voltage-to-Magnetic field calibration}
The  transduction between the applied voltage and the generated magnetic field at the vapour cell position is determined using the spin ensemble itself as a calibrated magnetometer. A small DC voltage $V_{DC}$ is applied to the RF coil to generate a transverse magnetic field $B_{y}$, producing a measurable shift in the Larmor frequency. The Larmor frequency in the presence of bias fields $B_{0}$ and $B_{y}$ is
\begin{equation}
    \Omega_{S} = g_{\mathrm{B}}\sqrt{B_{0}^{2}+ B_{y}^{2}},
\label{eq:gyromagnetic}
\end{equation}
where $g_{\mathrm{B}}/2\pi$ is the gyromagnetic ratio ($\sim 3.5\,  \mathrm{Hz}/\mathrm{nT}$). For $|B_{y}/B_{0}|\ll\,1 $ ($\leq 0.8 \%$ in Extended Data Fig.~\ref{fig:rf tone claibrations}a), expanding to the second order yields
\begin{equation}
    \Omega_{S}(V_{DC}) \approx g_{\mathrm{B}} B_{0}\left(1+\frac{1}{2}\frac{B_{y}^{2}}{B_{0}^{2}}\right).
\end{equation}
Since $B_{y} = \alpha_{B}V_{DC}$, fitting the quadratic dependence of $\Omega_{S}(V_{DC})$ on the applied DC voltage allows the extraction of the magnetic transduction coefficient $\alpha_{B}$ (Extended Data Fig.~\ref{fig:rf tone claibrations}b). This procedure provides the value $\alpha_{B} \approx$ 2.84\,$\mathrm{pT}/\mathrm{\upmu V}$ at the vapour-cell location.
To ensure that this DC field calibration remains valid at the measurement sideband frequencies, the RF coil transfer function was independently characterized. The RF coil ($R \approx5 \,\Omega$, $L\approx42\,\mathrm{\upmu H}$, $C \approx100 \,\mathrm{pF}$) is strongly damped with a 500\,$\Omega$ series resistor, resulting in a flat response from DC up to the 60 kHz operating Larmor frequency.

Having established the transduction coefficient $\alpha_{B}$, we can now determine the amplitude of the applied magnetic RF signals. The weak multi-tone RF fields are applied within $\pm5$\,kHz of the Larmor frequency to probe the magnetometer response without modifying the intrinsic spin susceptibility. The tones are spaced more densely near resonance to improve the spectral resolution there. The mean RF tone amplitude is calibrated from 7 independent 50-second traces processed with the same Welch parameters, from which a constant-amplitude fit to 37 RF tones yields $B_{\mathrm{RF}}= 127.0\pm 2.2\,\mathrm{fT}$ (1$\sigma$ combined uncertainty) using the calibrated transduction coefficient $\alpha_B$ .

\subsection{Magnetometer transfer function}

The magnetic-field calibration relies solely on the gyromagnetic relation~\eqref{eq:gyromagnetic} between the applied field and the atomic Larmor precession at $\theta_\mathrm{mag} = 0^{\circ}$ and is therefore independent of the chosen optical measurement strength $\Gamma_{S}$. In contrast, the magnetometer transfer function $\sqrt{\eta^{\mathrm{out}}_{\mathrm{mag}}\Gamma_{S}A_{\mathrm{B}}}\rho_{S}(\Omega){\cos}(\theta_{\mathrm{mag}})$ in the variational-readout and hybrid configurations depends on the optical readout settings and is therefore determined separately for each measurement configuration, by fitting the power spectral density of the recorded spin-noise spectra with and without RF-field excitation(Fig.~\ref{fig:Extended_fig_2}). The measured spin-noise amplitude spectra are then normalized by the atomic transfer function and referenced to the calibrated RF tone amplitude to reconstruct the noise-equivalent magnetic-field sensitivity spectra. Under identical experimental conditions, the transduction factor at acoustic frequencies is obtained by rescaling the calibrated value at 59\,kHz according to the readout rate ($\propto \Gamma_{S}^{10 \text{kHz}}/ \Gamma_{S}^{58 \text{kHz}}$), thereby accounting for the variation in the collective spin length.

\section{Benchmarking against the standard quantum limit}
To establish that our magnetometer is operating near the quantum-limited regime, we compare its sensitivity to the standard quantum limit (SQL). 
The sensitivity spectra are obtained from measured spin-noise spectra using system parameters extracted via an independent, cross-validated calibration procedure. Rather than assuming a power-independent oscillator decay rate $\gamma_S$ as in cavity optomechanics \cite{mason2019continuous, yu2020quantum}, we explicitly account for the power-dependence (broadening) of $\gamma_S$ and the atomic transduction factor $A_\mathrm{B}$ by analyzing the spin-noise spectra both with and without RF excitation. 

The multi-step calibration proceeds as follows. First, we extract the broadband decoherence rate $\gamma_{bb}$ and broadband measurement rate $\Gamma_{bb}$ by fitting the broadband spin noise spectra. The atomic Larmor frequency $\Omega_{S}$ and intrinsic spin decoherence rate $\gamma_{S}$ are then obtained by fitting the narrowband spin noise with a Lorentzian function. The measurement readout rate $\Gamma_{S}$ and thermal occupation $n_{S}$ are subsequently extracted by fitting the spin-noise spectra with the theory model, and the results are cross-validated against the observed ponderomotive squeezing level. Finally, the atomic transduction factor $A_\mathrm{B}$ is determined by fitting the spin-noise response with the full model under weak RF excitation while keeping all other parameters fixed. From these global fits, we deduce a consistent thermal spin occupation number of $n_{S} = 1.8$ across the entire power range.

Crucially, our analysis reveals that the atomic transduction factor $A_{\mathrm{B}}$ gradually degrades beyond a certain probe power threshold, implying a reduction in the collective spin length. Consequently, while the readout rate continues to scale up to a probe power of $\sim 3$\,mW, spin-length degradation prevents further improvements in peak sensitivity or detection bandwidth. This physical limitation highlights a bottleneck in classical power scaling and provides a strong direct motivation for non-classical approaches such as squeezed/entangled-light injection.

The probe power of 10\,$\upmu$W yields a maximum magnetic-field sensitivity of approximately 15\,fT/$\sqrt{\mathrm{Hz}}$. By varying the probe power, we map out the quantum sensing trade-off between measurement imprecision and quantum backaction. Around the analysis frequency detuning $(\Omega-\Omega_\mathrm{S})/2\pi$, the total noise is minimized at an optimal probe power, denoted by the coloured circles in Extended Data Fig.~\ref{fig:SQL validation}c, highlighting the frequency bands where a given power outperforms all others. The envelope of these power-optimized minima approaches the predicted SQL most closely at detunings of $\approx 1$--5\,kHz, confirming near-quantum-limited operation in this band.
We note that accurately modelling the high-frequency side of the spin-noise spectrum is inherently complicated by residual magnetic-field inhomogeneities and spectral contributions from the adjacent Zeeman manifolds near the 170\,kHz Larmor frequency. To ensure the reliability of our SQL validation, we enforced a calibration threshold: the residuals between the experimental data and the theoretical models for both the spin noise and RF-driven spin spectra were verified to remain within $\pm10\,\%$ across the off-resonant analysis window before reconstructing the final sensitivity curves. During the SQL validation, a different Cs cell of identical geometry was used at the same operating temperature, because the atomic density of the original cell had degraded after the sensitivity measurements discussed in the main text and no longer provided the density required for a reliable SQL comparison. The aim of this calibration is to benchmark the theoretical SQL model against experimental data rather than to characterize a single cell — once this agreement is established, the SQL applies to any cell through its own parameters, so the choice of calibration cell does not affect the results.
\section{Variational readout}
With the system calibrated against the SQL, we now analyze the spin-noise spectra at different variational readout angles to calibrate the atomic transfer function and quantify the underlying quantum noise contributions. The spin-noise spectra are first measured without RF excitation (color curves in Extended Data Fig.~\ref{fig:Extended_fig_2}). These spectra are fitted with all model parameters shared across the datasets except the detection angle $\theta_{\mathrm{mag}}$, which is fitted independently for each. The extracted parameters are cross-validated by their consistency with the observed maximum ponderomotive squeezing \cite{Jia2023}. RF-driven spin-noise spectra are recorded under identical conditions and fitted using the full model including the driven response, with the magnetic transduction factor $A_{B}$ as the only additional free parameter. The resulting fit captures the response amplitude at the multi-tone RF frequencies (gray circles), confirming the calibrated transfer function. A small systematic deviation is observed within $\pm 500$\,Hz of the Larmor resonance, where the Lorentzian susceptibility model accurately captures the peak response but does not fully reproduce the off-resonant lineshape. This discrepancy is reduced by a Voigt-profile model, suggesting additional inhomogeneous broadening not further explored here. 
Using the extracted transfer function, the noise-equivalent magnetic-field sensitivity is reconstructed for each variational readout angle, as presented in the main text. The SQL and frequency-optimized sensitivity curves are obtained using a calibrated transduction factor $A_\mathrm{B}$ averaged over five detection angles.
A slight offset between the frequency of maximum ponderomotive squeezing and the optimal sensitivity (marked by vertical dashed lines) is observed, consistent with theoretical predictions for variational readout in continuous linear systems, as discussed in the context of gravitational-wave detectors \cite{Kimble2001}.

\section{EPR conditioning}
The vacuum mode in the probe orthogonal polarization is replaced by one mode of an EPR-entangled source. The other mode bypasses the atoms and is detected as a reference channel, while the atomic arm is measured at a phase quadrature $\theta_{\mathrm{mag}} = 0^{\circ}$. The conditional probe quadrature noise is constructed by optimally combining two photocurrents using a Wiener filter (see Supplementary Materials for details) \cite{gould2021optimal,novikov2025hybrid}, without involving atoms, yielding $\sim 4$\,dB noise reduction relative to the single-arm shot-noise level (Extended Data Fig.~\ref{fig:extend_figure_4}a).

Conditioned spin-noise spectra are obtained by scanning the EPR readout phase $\phi_{\mathrm{EPR}}$ and exploiting the cross-correlations between the probe and reference channels. Up to $\sim 3$\,dB noise reduction relative to the vacuum-driven spin-noise is observed (Extended Data Fig.~\ref{fig:extend_figure_4}b). The spectra are fitted using the full spin-noise model with a common set of parameters, while $\phi_{\mathrm{EPR}}$ is treated as an independent variable. The conditional noise reduction is mapped as a function of sideband frequency and $\phi_{\mathrm{EPR}}$ (2D contour), from which the optimal frequency-dependent conditioning angle, $\arctan(\Gamma_{S}\chi_{S}(\Omega))$, is identified (dash-dotted line) \cite{novikov2025hybrid}. 
Under identical experimental conditions, the transduction factor at acoustic frequencies is obtained by rescaling the calibrated value at 58 kHz according to the extracted measurement readout rate $\Gamma_{S}$, reflecting an increased effective collective spin length inferred from the spin-noise spectra. Using this calibrated transfer function, the conditioned spectra are converted into magnetic-field sensitivity. The optimal conditioning phase coincides with the angle that minimizes the reconstructed noise-equivalent sensitivity. The same transduction factor is used in the SQL and optimal field sensitivity reconstruction at 10\,kHz.

\section{Hybrid configuration}
We combine variational readout and EPR conditioning to simultaneously exploit correlations between imprecision and backaction noise, as well as between the probe and the entangled reference channel. In this hybrid configuration, variational readout first generates ponderomotive squeezing of the probe field, after which the remaining quantum noise is further suppressed using correlations with the reference channel.
The atomic detection quadrature is fixed at an intermediate angle, $\theta_{\mathrm{mag}} = -55^\circ$, inferred from a fit to the variational readout spin-noise spectra driven by EPR fluctuations. Under the same operating conditions, spin-noise spectra are recorded and fitted using the hybrid model (see Supplementary Materials for details), allowing only the EPR conditioning phase to vary while holding all remaining parameters at the values determined in the EPR conditioning characterization. The vacuum-noise-driven spin readout reference is reconstructed including the signal attenuation associated with the rotated detection quadrature $\theta_{\mathrm{mag}}$, as described for variational readout.
At this detection angle, up to $\sim 5$\,dB noise reduction relative to the reconstructed spin-noise reference is observed below the Larmor frequency. This enhancement is visualized in the 2D map of conditional noise reduction (Extended Data Fig.~\ref{fig:extend_figure_4}c), where the optimal conditioning phase predicted by the effective atomic susceptibility in our previous work \cite{Zeuthen2019, novikov2025hybrid}, reproduces the minima across the full frequency range.
Using the rescaled transduction factor that accounts for the detection angle $\theta_{\mathrm{mag}}$ given by expression (\ref{eqM:delta-B_y}), the spectra are converted into magnetic-field sensitivity. The same factor is used to reconstruct the frequency-optimized sensitivity at fixed variational readout angle, while the SQL reference is kept identical to that used in the EPR conditioning configuration. Based on the hybrid configuration model and extracted parameters, the optimal sensitivity is further obtained by numerically optimizing both the variational readout angle $\theta_{\mathrm{mag}}$ and the EPR conditioning phase $\phi_{\mathrm{EPR}}$ at each sideband frequency (see Supplementary Material for details).
\section{Uncertainty analysis and model uncertainty}
During preprocessing, both the multi-tone RF and spin-dynamic signals are recorded for 50\,s at 1\,MHz sampling rate and processed using Welch method (Hann window, 75$\%$ segment overlap, and 2\,Hz resolution). This yields a relative uncertainty of $\approx 7\%$ for the power spectral density ($1\sigma$), corresponding to $\approx 3.5\%$ uncertainty in the amplitude spectrum, with effective degrees of freedom
$\approx 420$~\cite{solomon1991psd}. 

These uncertainties weight the joint least-squares fit of 18 spin-noise spectra to the conditional-squeezing model, with shared physics parameters and dataset-specific conditioning phases, to extract the joint parameter covariance matrix.
Uncertainties on the sub-SQL performance are propagated via a parametric Monte Carlo bootstrap ($N = 10^{4}$): parameters are drawn from a multivariate Gaussian by the best-fit parameters and their fit covariance matrix.  The sub-SQL margin is evaluated at the frequency of maximum quantum noise reduction predicted by the joint fit and the reported value and $1\sigma$ uncertainty are the mean and standard deviation of the resulting bootstrap distribution. During the sub-SQL analysis, the transduction coefficient $A_\mathrm{B}$ and the RF field amplitude cancel in the ratio $S_{\mathrm{meas}}/S_{\mathrm{SQL}}$, and therefore do not contribute to the sub-SQL uncertainty budget.

The absolute magnetic field sensitivity is reconstructed using the calibrated RF amplitude $B_{\mathrm{RF}} = 127.0 \pm 2.2\,\text{fT}$ and the model transduction extracted from the spin-noise fit. Combining the uncertainty (1.7$\%$) of the RF calibration tones with the model-data agreement in the Extended Data Fig.~\ref{fig:Extended_fig_2} of $\pm 10\%$ on both the undriven and RF-driven spin noise power spectra, error propagation yields a combined relative uncertainty of $\approx 7\%$ ($1\sigma$) on the reconstructed magnetic field sensitivity. The reported peak sensitivity of $47\, \text{fT}/\sqrt{\mathrm{Hz}}$ thus carries an uncertainty of $\pm 3\,\text{fT}/\sqrt{\mathrm{Hz}}$. For a conservative estimate on the uncertainty of the field sensitivity, one can adopt the $\pm 10\%$ model-data agreement rather than pursuing a tighter bound. 
\FloatBarrier
\bibliographystyle{naturemag_format_title}
\bibliography{references}

\begin{thebibliography}{14}%
\makeatletter
\providecommand \@ifxundefined [1]{%
 \@ifx{#1\undefined}
}%
\providecommand \@ifnum [1]{%
 \ifnum #1\expandafter \@firstoftwo
 \else \expandafter \@secondoftwo
 \fi
}%
\providecommand \@ifx [1]{%
 \ifx #1\expandafter \@firstoftwo
 \else \expandafter \@secondoftwo
 \fi
}%
\providecommand \natexlab [1]{#1}%
\providecommand \enquote  [1]{``#1''}%
\providecommand \bibnamefont  [1]{#1}%
\providecommand \bibfnamefont [1]{#1}%
\providecommand \citenamefont [1]{#1}%
\providecommand \href@noop [0]{\@secondoftwo}%
\providecommand \href [0]{\begingroup \@sanitize@url \@href}%
\providecommand \@href[1]{\@@startlink{#1}\@@href}%
\providecommand \@@href[1]{\endgroup#1\@@endlink}%
\providecommand \@sanitize@url [0]{\catcode `\\12\catcode `\$12\catcode `\&12\catcode `\#12\catcode `\^12\catcode `\_12\catcode `\%12\relax}%
\providecommand \@@startlink[1]{}%
\providecommand \@@endlink[0]{}%
\providecommand \url  [0]{\begingroup\@sanitize@url \@url }%
\providecommand \@url [1]{\endgroup\@href {#1}{\urlprefix }}%
\providecommand \urlprefix  [0]{URL }%
\providecommand \Eprint [0]{\href }%
\providecommand \doibase [0]{https://doi.org/}%
\providecommand \selectlanguage [0]{\@gobble}%
\providecommand \bibinfo  [0]{\@secondoftwo}%
\providecommand \bibfield  [0]{\@secondoftwo}%
\providecommand \translation [1]{[#1]}%
\providecommand \BibitemOpen [0]{}%
\providecommand \bibitemStop [0]{}%
\providecommand \bibitemNoStop [0]{.\EOS\space}%
\providecommand \EOS [0]{\spacefactor3000\relax}%
\providecommand \BibitemShut  [1]{\csname bibitem#1\endcsname}%
\let\auto@bib@innerbib\@empty
\bibitem [{\citenamefont {Steck}(2003)}]{steck2003cesium}%
  \BibitemOpen
  \bibfield  {author} {\bibinfo {author} {\bibfnamefont {D.~A.}\ \bibnamefont {Steck}},\ }\bibfield  {title} {\bibinfo {title} {{Cesium D line data}},\ }\href@noop {} {\  (\bibinfo {year} {2003})}\BibitemShut {NoStop}%
\bibitem [{\citenamefont {Thomas}\ \emph {et~al.}(2021)\citenamefont {Thomas}, \citenamefont {Parniak}, \citenamefont {\O{}stfeldt} \emph {et~al.}}]{Thomas2021}%
  \BibitemOpen
  \bibfield  {author} {\bibinfo {author} {\bibfnamefont {R.~A.}\ \bibnamefont {Thomas}}, \bibinfo {author} {\bibfnamefont {M.}~\bibnamefont {Parniak}}, \bibinfo {author} {\bibfnamefont {C.}~\bibnamefont {\O{}stfeldt}}, \emph {et~al.},\ }\bibfield  {title} {\bibinfo {title} {Entanglement between distant macroscopic mechanical and spin systems},\ }\href {https://doi.org/10.1038/s41567-020-1031-5} {\bibfield  {journal} {\bibinfo  {journal} {Nat. Phys.}\ }\textbf {\bibinfo {volume} {17}},\ \bibinfo {pages} {228–233} (\bibinfo {year} {2021})}\BibitemShut {NoStop}%
\bibitem [{\citenamefont {Novikov}\ \emph {et~al.}(2025)\citenamefont {Novikov}, \citenamefont {Jia}, \citenamefont {Brasil}, \citenamefont {Grimaldi}, \citenamefont {Bocoum}, \citenamefont {Balabas}, \citenamefont {M{\"u}ller}, \citenamefont {Zeuthen},\ and\ \citenamefont {Polzik}}]{novikov2025hybrid}%
  \BibitemOpen
  \bibfield  {author} {\bibinfo {author} {\bibfnamefont {V.}~\bibnamefont {Novikov}}, \bibinfo {author} {\bibfnamefont {J.}~\bibnamefont {Jia}}, \bibinfo {author} {\bibfnamefont {T.~B.}\ \bibnamefont {Brasil}}, \bibinfo {author} {\bibfnamefont {A.}~\bibnamefont {Grimaldi}}, \bibinfo {author} {\bibfnamefont {M.}~\bibnamefont {Bocoum}}, \bibinfo {author} {\bibfnamefont {M.}~\bibnamefont {Balabas}}, \bibinfo {author} {\bibfnamefont {J.~H.}\ \bibnamefont {M{\"u}ller}}, \bibinfo {author} {\bibfnamefont {E.}~\bibnamefont {Zeuthen}},\ and\ \bibinfo {author} {\bibfnamefont {E.~S.}\ \bibnamefont {Polzik}},\ }\bibfield  {title} {\bibinfo {title} {Hybrid quantum network for sensing in the acoustic frequency range},\ }\href@noop {} {\bibfield  {journal} {\bibinfo  {journal} {Nature}\ }\textbf {\bibinfo {volume} {643}},\ \bibinfo {pages} {955} (\bibinfo {year} {2025})}\BibitemShut {NoStop}%
\bibitem [{\citenamefont {Julsgaard}(2003)}]{Julsgaard2003thesis}%
  \BibitemOpen
  \bibfield  {author} {\bibinfo {author} {\bibfnamefont {B.}~\bibnamefont {Julsgaard}},\ }\emph {\bibinfo {title} {Entanglement and Quantum Interactions with Macroscopic Gas Samples}},\ \href {https://phys.au.dk/fileadmin/site_files/publikationer/phd/Brian_Julsgaard.pdf} {\bibinfo {type} {Ph.d. thesis}},\ \bibinfo  {school} {University of Aarhus} (\bibinfo {year} {2003})\BibitemShut {NoStop}%
\bibitem [{\citenamefont {Danilishin}\ and\ \citenamefont {Khalili}(2012)}]{Danilishin2012}%
  \BibitemOpen
  \bibfield  {author} {\bibinfo {author} {\bibfnamefont {S.~L.}\ \bibnamefont {Danilishin}}\ and\ \bibinfo {author} {\bibfnamefont {F.~Y.}\ \bibnamefont {Khalili}},\ }\bibfield  {title} {\bibinfo {title} {Quantum measurement theory in gravitational-wave detectors},\ }\bibfield  {journal} {\bibinfo  {journal} {Living Rev. Relativ.}\ }\textbf {\bibinfo {volume} {15}},\ \href {https://doi.org/10.12942/lrr-2012-5} {10.12942/lrr-2012-5} (\bibinfo {year} {2012})\BibitemShut {NoStop}%
\bibitem [{\citenamefont {Danilishin}\ \emph {et~al.}(2019)\citenamefont {Danilishin}, \citenamefont {Khalili},\ and\ \citenamefont {Miao}}]{Danilishin2019}%
  \BibitemOpen
  \bibfield  {author} {\bibinfo {author} {\bibfnamefont {S.~L.}\ \bibnamefont {Danilishin}}, \bibinfo {author} {\bibfnamefont {F.~Y.}\ \bibnamefont {Khalili}},\ and\ \bibinfo {author} {\bibfnamefont {H.}~\bibnamefont {Miao}},\ }\bibfield  {title} {\bibinfo {title} {Advanced quantum techniques for future gravitational-wave detectors},\ }\bibfield  {journal} {\bibinfo  {journal} {Living Rev. Relativ.}\ }\textbf {\bibinfo {volume} {22}},\ \href {https://doi.org/10.1007/s41114-019-0018-y} {10.1007/s41114-019-0018-y} (\bibinfo {year} {2019})\BibitemShut {NoStop}%
\bibitem [{\citenamefont {Braginsky}(1967)}]{braginsky1967classical}%
  \BibitemOpen
  \bibfield  {author} {\bibinfo {author} {\bibfnamefont {V.}~\bibnamefont {Braginsky}},\ }\bibfield  {title} {\bibinfo {title} {Classical and quantum restrictions on the detection of weak disturbances of a macroscopic oscillator},\ }\href@noop {} {\bibfield  {journal} {\bibinfo  {journal} {Zh. Eksp. Teor. Fiz}\ }\textbf {\bibinfo {volume} {53}},\ \bibinfo {pages} {1434} (\bibinfo {year} {1967})}\BibitemShut {NoStop}%
\bibitem [{\citenamefont {Braginsky}\ and\ \citenamefont {Khalili}(1995)}]{braginsky1995quantum}%
  \BibitemOpen
  \bibfield  {author} {\bibinfo {author} {\bibfnamefont {V.~B.}\ \bibnamefont {Braginsky}}\ and\ \bibinfo {author} {\bibfnamefont {F.~Y.}\ \bibnamefont {Khalili}},\ }\href@noop {} {\emph {\bibinfo {title} {Quantum measurement}}}\ (\bibinfo  {publisher} {Cambridge University Press},\ \bibinfo {year} {1995})\BibitemShut {NoStop}%
\bibitem [{\citenamefont {Khalili}\ and\ \citenamefont {Zeuthen}(2021)}]{khalili2021quantum}%
  \BibitemOpen
  \bibfield  {author} {\bibinfo {author} {\bibfnamefont {F.~Y.}\ \bibnamefont {Khalili}}\ and\ \bibinfo {author} {\bibfnamefont {E.}~\bibnamefont {Zeuthen}},\ }\bibfield  {title} {\bibinfo {title} {Quantum limits for stationary force sensing},\ }\href@noop {} {\bibfield  {journal} {\bibinfo  {journal} {Phys. Rev. A}\ }\textbf {\bibinfo {volume} {103}},\ \bibinfo {pages} {043721} (\bibinfo {year} {2021})}\BibitemShut {NoStop}%
\bibitem [{\citenamefont {Yu}\ \emph {et~al.}(2020)\citenamefont {Yu}, \citenamefont {McCuller}, \citenamefont {Tse}, \citenamefont {Kijbunchoo}, \citenamefont {Barsotti},\ and\ \citenamefont {Mavalvala}}]{yu2020quantum}%
  \BibitemOpen
  \bibfield  {author} {\bibinfo {author} {\bibfnamefont {H.}~\bibnamefont {Yu}}, \bibinfo {author} {\bibfnamefont {L.}~\bibnamefont {McCuller}}, \bibinfo {author} {\bibfnamefont {M.}~\bibnamefont {Tse}}, \bibinfo {author} {\bibfnamefont {N.}~\bibnamefont {Kijbunchoo}}, \bibinfo {author} {\bibfnamefont {L.}~\bibnamefont {Barsotti}},\ and\ \bibinfo {author} {\bibfnamefont {N.}~\bibnamefont {Mavalvala}},\ }\bibfield  {title} {\bibinfo {title} {Quantum correlations between light and the kilogram-mass mirrors of ligo},\ }\href@noop {} {\bibfield  {journal} {\bibinfo  {journal} {Nature}\ }\textbf {\bibinfo {volume} {583}},\ \bibinfo {pages} {43} (\bibinfo {year} {2020})}\BibitemShut {NoStop}%
\bibitem [{\citenamefont {Brown}\ and\ \citenamefont {Hwang}(1997)}]{brown1997introduction}%
  \BibitemOpen
  \bibfield  {author} {\bibinfo {author} {\bibfnamefont {R.~G.}\ \bibnamefont {Brown}}\ and\ \bibinfo {author} {\bibfnamefont {P.~Y.}\ \bibnamefont {Hwang}},\ }\bibfield  {title} {\bibinfo {title} {Introduction to random signals and applied kalman filtering: with matlab exercises and solutions},\ }\href@noop {} {\bibfield  {journal} {\bibinfo  {journal} {Introduction to random signals and applied Kalman filtering: with MATLAB exercises and solutions}\ } (\bibinfo {year} {1997})}\BibitemShut {NoStop}%
\bibitem [{\citenamefont {Gould}\ \emph {et~al.}(2021)\citenamefont {Gould}, \citenamefont {Yap}, \citenamefont {Adya}, \citenamefont {Slagmolen}, \citenamefont {Ward},\ and\ \citenamefont {McClelland}}]{gould2021optimal}%
  \BibitemOpen
  \bibfield  {author} {\bibinfo {author} {\bibfnamefont {D.~W.}\ \bibnamefont {Gould}}, \bibinfo {author} {\bibfnamefont {M.~J.}\ \bibnamefont {Yap}}, \bibinfo {author} {\bibfnamefont {V.~B.}\ \bibnamefont {Adya}}, \bibinfo {author} {\bibfnamefont {B.~J.}\ \bibnamefont {Slagmolen}}, \bibinfo {author} {\bibfnamefont {R.~L.}\ \bibnamefont {Ward}},\ and\ \bibinfo {author} {\bibfnamefont {D.~E.}\ \bibnamefont {McClelland}},\ }\bibfield  {title} {\bibinfo {title} {Optimal quantum noise cancellation with an entangled witness channel},\ }\href@noop {} {\bibfield  {journal} {\bibinfo  {journal} {Phys. Rev. Research}\ }\textbf {\bibinfo {volume} {3}},\ \bibinfo {pages} {043079} (\bibinfo {year} {2021})}\BibitemShut {NoStop}%
\bibitem [{\citenamefont {Duan}\ \emph {et~al.}(2000)\citenamefont {Duan}, \citenamefont {Giedke}, \citenamefont {Cirac},\ and\ \citenamefont {Zoller}}]{duan2000entanglement}%
  \BibitemOpen
  \bibfield  {author} {\bibinfo {author} {\bibfnamefont {L.-M.}\ \bibnamefont {Duan}}, \bibinfo {author} {\bibfnamefont {G.}~\bibnamefont {Giedke}}, \bibinfo {author} {\bibfnamefont {J.~I.}\ \bibnamefont {Cirac}},\ and\ \bibinfo {author} {\bibfnamefont {P.}~\bibnamefont {Zoller}},\ }\bibfield  {title} {\bibinfo {title} {Entanglement purification of gaussian continuous variable quantum states},\ }\href@noop {} {\bibfield  {journal} {\bibinfo  {journal} {Phys. Rev. Lett.}\ }\textbf {\bibinfo {volume} {84}},\ \bibinfo {pages} {4002} (\bibinfo {year} {2000})}\BibitemShut {NoStop}%
\bibitem [{\citenamefont {Andalkar}\ and\ \citenamefont {Warrington}(2002)}]{voigt}%
  \BibitemOpen
  \bibfield  {author} {\bibinfo {author} {\bibfnamefont {A.}~\bibnamefont {Andalkar}}\ and\ \bibinfo {author} {\bibfnamefont {R.~B.}\ \bibnamefont {Warrington}},\ }\bibfield  {title} {\bibinfo {title} {High-resolution measurement of the pressure broadening and shift of the cs $d1$ and $d2$ lines by ${\mathrm{n}}_{2}$ and he buffer gases},\ }\href {https://doi.org/10.1103/PhysRevA.65.032708} {\bibfield  {journal} {\bibinfo  {journal} {Phys. Rev. A}\ }\textbf {\bibinfo {volume} {65}},\ \bibinfo {pages} {032708} (\bibinfo {year} {2002})}\BibitemShut {NoStop}%
\end{thebibliography}%


\begin{thebibliography}{10}
\expandafter\ifx\csname url\endcsname\relax
  \def\url#1{\texttt{#1}}\fi
\expandafter\ifx\csname urlprefix\endcsname\relax\def\urlprefix{URL }\fi
\providecommand{\bibinfo}[2]{#2}
\providecommand{\eprint}[2][]{\url{#2}}

\bibitem{clerk2010introduction}
\bibinfo{author}{Clerk, A.~A.}, \bibinfo{author}{Devoret, M.~H.}, \bibinfo{author}{Girvin, S.~M.}, \bibinfo{author}{Marquardt, F.} \& \bibinfo{author}{Schoelkopf, R.~J.}
\newblock Introduction to quantum noise, measurement, and amplification.
\newblock \emph{\bibinfo{journal}{Rev. Mod. Phys.}} \textbf{\bibinfo{volume}{82}}, \bibinfo{pages}{1155--1208} (\bibinfo{year}{2010}).

\bibitem{braginsky1995quantum}
\bibinfo{author}{Braginsky, V.~B.} \& \bibinfo{author}{Khalili, F.~Y.}
\newblock \emph{\bibinfo{title}{Quantum measurement}} (\bibinfo{publisher}{Cambridge University Press}, \bibinfo{year}{1995}).

\bibitem{markus}
\bibinfo{author}{Aspelmeyer, M.}, \bibinfo{author}{Kippenberg, T.~J.} \& \bibinfo{author}{Marquardt, F.}
\newblock Cavity optomechanics.
\newblock \emph{\bibinfo{journal}{Rev. Mod. Phys.}} \textbf{\bibinfo{volume}{86}}, \bibinfo{pages}{1391--1452}.
\newblock \urlprefix\url{https://link.aps.org/doi/10.1103/RevModPhys.86.1391}.

\bibitem{braginsky1967classical}
\bibinfo{author}{Braginsky, V.}
\newblock Classical and quantum restrictions on the detection of weak disturbances of a macroscopic oscillator.
\newblock \emph{\bibinfo{journal}{Zh. Eksp. Teor. Fiz}} \textbf{\bibinfo{volume}{53}}, \bibinfo{pages}{1434--1441} (\bibinfo{year}{1967}).

\bibitem{khalili2021quantum}
\bibinfo{author}{Khalili, F.~Y.} \& \bibinfo{author}{Zeuthen, E.}
\newblock Quantum limits for stationary force sensing.
\newblock \emph{\bibinfo{journal}{Phys. Rev. A}} \textbf{\bibinfo{volume}{103}}, \bibinfo{pages}{043721} (\bibinfo{year}{2021}).

\bibitem{Danilishin2012}
\bibinfo{author}{Danilishin, S.~L.} \& \bibinfo{author}{Khalili, F.~Y.}
\newblock Quantum Measurement Theory in Gravitational-Wave Detectors.
\newblock \emph{\bibinfo{journal}{Living Rev. Relativ.}} \textbf{\bibinfo{volume}{15}} (\bibinfo{year}{2012}).
\newblock \urlprefix\url{http://dx.doi.org/10.12942/lrr-2012-5}.

\bibitem{budker2007optical}
\bibinfo{author}{Budker, D.} \& \bibinfo{author}{Romalis, M.}
\newblock Optical magnetometry.
\newblock \emph{\bibinfo{journal}{Nat. Phys.}} \textbf{\bibinfo{volume}{3}}, \bibinfo{pages}{227--234} (\bibinfo{year}{2007}).

\bibitem{Budker_Jackson_Kimball_2013}
\bibinfo{author}{Budker, D.} \& \bibinfo{author}{Kimball, D. F.~J.}
\newblock \emph{\bibinfo{title}{Optical Magnetometry}} (\bibinfo{publisher}{Cambridge University Press}, \bibinfo{year}{2013}).

\bibitem{wasilewski2010quantum}
\bibinfo{author}{Wasilewski, W.} \emph{et~al.}
\newblock Quantum noise limited and entanglement-assisted magnetometry.
\newblock \emph{\bibinfo{journal}{Phys. Rev. Lett.}} \textbf{\bibinfo{volume}{104}}, \bibinfo{pages}{133601} (\bibinfo{year}{2010}).

\bibitem{fleischhauer2000quantum}
\bibinfo{author}{Fleischhauer, M.}, \bibinfo{author}{Matsko, A.} \& \bibinfo{author}{Scully, M.}
\newblock Quantum limit of optical magnetometry in the presence of ac Stark shifts.
\newblock \emph{\bibinfo{journal}{Phys. Rev. A}} \textbf{\bibinfo{volume}{62}}, \bibinfo{pages}{013808} (\bibinfo{year}{2000}).

\bibitem{vasilyev2012quantum}
\bibinfo{author}{Vasilyev, D.~V.}, \bibinfo{author}{Hammerer, K.}, \bibinfo{author}{Korolev, N.} \& \bibinfo{author}{S{\o}rensen, A.~S.}
\newblock Quantum noise for faraday light--matter interfaces.
\newblock \emph{\bibinfo{journal}{J. Phys. B}} \textbf{\bibinfo{volume}{45}}, \bibinfo{pages}{124007} (\bibinfo{year}{2012}).

\bibitem{Braginskii1996}
\bibinfo{author}{Braginsky, V.} \& \bibinfo{author}{Khalili, F.}
\newblock {Quantum nondemolition measurements: the route from toys to tools}.
\newblock \emph{\bibinfo{journal}{Rev. Mod. Phys.}} \textbf{\bibinfo{volume}{68}} (\bibinfo{year}{1996}).
\newblock \urlprefix\url{https://journals.aps.org/rmp/abstract/10.1103/RevModPhys.68.1}.

\bibitem{vasilakis2015generation}
\bibinfo{author}{Vasilakis, G.} \emph{et~al.}
\newblock Generation of a squeezed state of an oscillator by stroboscopic back-action-evading measurement.
\newblock \emph{\bibinfo{journal}{Nat. Phys.}} \textbf{\bibinfo{volume}{11}}, \bibinfo{pages}{389--392} (\bibinfo{year}{2015}).

\bibitem{colangelo2017simultaneous}
\bibinfo{author}{Colangelo, G.}, \bibinfo{author}{Ciurana, F.~M.}, \bibinfo{author}{Bianchet, L.~C.}, \bibinfo{author}{Sewell, R.~J.} \& \bibinfo{author}{Mitchell, M.~W.}
\newblock Simultaneous tracking of spin angle and amplitude beyond classical limits.
\newblock \emph{\bibinfo{journal}{Nature}} \textbf{\bibinfo{volume}{543}}, \bibinfo{pages}{525--528} (\bibinfo{year}{2017}).

\bibitem{troullinou2021squeezed}
\bibinfo{author}{Troullinou, C.}, \bibinfo{author}{Jim{\'e}nez-Mart{\'\i}nez, R.}, \bibinfo{author}{Kong, J.}, \bibinfo{author}{Lucivero, V.} \& \bibinfo{author}{Mitchell, M.}
\newblock Squeezed-light enhancement and backaction evasion in a high sensitivity optically pumped magnetometer.
\newblock \emph{\bibinfo{journal}{Phys. Rev. Lett.}} \textbf{\bibinfo{volume}{127}}, \bibinfo{pages}{193601} (\bibinfo{year}{2021}).

\bibitem{Polzik2015}
\bibinfo{author}{Polzik, E.~S.} \& \bibinfo{author}{Hammerer, K.}
\newblock Trajectories without quantum uncertainties.
\newblock \emph{\bibinfo{journal}{Annalen der Physik}} \textbf{\bibinfo{volume}{527}} (\bibinfo{year}{2015}).
\newblock \urlprefix\url{http://dx.doi.org/10.1002/andp.201400099}.

\bibitem{Muller2017}
\bibinfo{author}{M\o{}ller, C.~B.}, \bibinfo{author}{Thomas, R.~A.}, \bibinfo{author}{Vasilakis, G.} \emph{et~al.}
\newblock Quantum back-action-evading measurement of motion in a negative mass reference frame.
\newblock \emph{\bibinfo{journal}{Nature}} \textbf{\bibinfo{volume}{547}}, \bibinfo{pages}{191–195} (\bibinfo{year}{2017}).
\newblock \urlprefix\url{http://dx.doi.org/10.1038/nature22980}.

\bibitem{kampel2017improving}
\bibinfo{author}{Kampel, N.} \emph{et~al.}
\newblock Improving broadband displacement detection with quantum correlations.
\newblock \emph{\bibinfo{journal}{Phys. Rev. X}} \textbf{\bibinfo{volume}{7}}, \bibinfo{pages}{021008} (\bibinfo{year}{2017}).

\bibitem{mason2019continuous}
\bibinfo{author}{Mason, D.}, \bibinfo{author}{Chen, J.}, \bibinfo{author}{Rossi, M.}, \bibinfo{author}{Tsaturyan, Y.} \& \bibinfo{author}{Schliesser, A.}
\newblock Continuous force and displacement measurement below the standard quantum limit.
\newblock \emph{\bibinfo{journal}{Nat. Phys.}} \textbf{\bibinfo{volume}{15}}, \bibinfo{pages}{745--749} (\bibinfo{year}{2019}).

\bibitem{baerentsen2024squeezed}
\bibinfo{author}{B{\ae}rentsen, C.} \emph{et~al.}
\newblock Squeezed light from an oscillator measured at the rate of oscillation.
\newblock \emph{\bibinfo{journal}{Nat. Commun.}} \textbf{\bibinfo{volume}{15}}, \bibinfo{pages}{4146} (\bibinfo{year}{2024}).

\bibitem{yap2020broadband}
\bibinfo{author}{Yap, M.~J.}, \bibinfo{author}{Cripe, J.}, \bibinfo{author}{Mansell, G.~L.} \emph{et~al.}
\newblock Broadband reduction of quantum radiation pressure noise via squeezed light injection.
\newblock \emph{\bibinfo{journal}{Nat. Photon.}} \textbf{\bibinfo{volume}{14}}, \bibinfo{pages}{19--23} (\bibinfo{year}{2020}).

\bibitem{Ganapathy2023}
\bibinfo{author}{Ganapathy, D.}, \bibinfo{author}{Jia, W.}, \bibinfo{author}{Nakano, M.} \emph{et~al.}
\newblock Broadband Quantum Enhancement of the LIGO Detectors with Frequency-Dependent Squeezing.
\newblock \emph{\bibinfo{journal}{Phys. Rev. X}} \textbf{\bibinfo{volume}{13}}, \bibinfo{pages}{041021} (\bibinfo{year}{2023}).
\newblock \urlprefix\url{https://link.aps.org/doi/10.1103/PhysRevX.13.041021}.

\bibitem{jia2024squeezing}
\bibinfo{author}{Jia, W.} \emph{et~al.}
\newblock Squeezing the quantum noise of a gravitational-wave detector below the standard quantum limit.
\newblock \emph{\bibinfo{journal}{Science}} \textbf{\bibinfo{volume}{385}}, \bibinfo{pages}{1318--1321} (\bibinfo{year}{2024}).

\bibitem{Danilishin2019}
\bibinfo{author}{Danilishin, S.~L.}, \bibinfo{author}{Khalili, F.~Y.} \& \bibinfo{author}{Miao, H.}
\newblock Advanced quantum techniques for future gravitational-wave detectors.
\newblock \emph{\bibinfo{journal}{Living Rev. Relativ.}} \textbf{\bibinfo{volume}{22}} (\bibinfo{year}{2019}).
\newblock \urlprefix\url{http://dx.doi.org/10.1007/s41114-019-0018-y}.

\bibitem{Aslam2023}
\bibinfo{author}{Aslam, N.}, \bibinfo{author}{Zhou, H.}, \bibinfo{author}{Urbach, E.~K.} \emph{et~al.}
\newblock Quantum sensors for biomedical applications.
\newblock \emph{\bibinfo{journal}{Nat. Rev. Phys.}} \textbf{\bibinfo{volume}{5}}, \bibinfo{pages}{157–169} (\bibinfo{year}{2023}).
\newblock \urlprefix\url{http://dx.doi.org/10.1038/s42254-023-00558-3}.

\bibitem{hamalainen1993magnetoencephalography}
\bibinfo{author}{H{\"a}m{\"a}l{\"a}inen, M.}, \bibinfo{author}{Hari, R.}, \bibinfo{author}{Ilmoniemi, R.~J.}, \bibinfo{author}{Knuutila, J.} \& \bibinfo{author}{Lounasmaa, O.~V.}
\newblock Magnetoencephalography—theory, instrumentation, and applications to noninvasive studies of the working human brain.
\newblock \emph{\bibinfo{journal}{Rev. Mod. Phys.}} \textbf{\bibinfo{volume}{65}}, \bibinfo{pages}{413} (\bibinfo{year}{1993}).

\bibitem{stuart1972earth}
\bibinfo{author}{Stuart, W.}
\newblock Earth's field magnetometry.
\newblock \emph{\bibinfo{journal}{Rep. Prog. Phys.}} \textbf{\bibinfo{volume}{35}}, \bibinfo{pages}{803--881} (\bibinfo{year}{1972}).

\bibitem{rikitake1968geomagnetism}
\bibinfo{author}{Rikitake, T.}
\newblock Geomagnetism and earthquake prediction.
\newblock \emph{\bibinfo{journal}{Tectonophysics}} \textbf{\bibinfo{volume}{6}}, \bibinfo{pages}{59--68} (\bibinfo{year}{1968}).

\bibitem{glenn2017micrometer}
\bibinfo{author}{Glenn, D.~R.} \emph{et~al.}
\newblock Micrometer-scale magnetic imaging of geological samples using a quantum diamond microscope.
\newblock \emph{\bibinfo{journal}{Geochem. Geophys. Geosyst.}} \textbf{\bibinfo{volume}{18}}, \bibinfo{pages}{3254--3267} (\bibinfo{year}{2017}).

\bibitem{hammerer2010quantum}
\bibinfo{author}{Hammerer, K.}, \bibinfo{author}{S{\o}rensen, A.~S.} \& \bibinfo{author}{Polzik, E.~S.}
\newblock Quantum interface between light and atomic ensembles.
\newblock \emph{\bibinfo{journal}{Rev. Mod. Phys.}} \textbf{\bibinfo{volume}{82}}, \bibinfo{pages}{1041--1093} (\bibinfo{year}{2010}).

\bibitem{geremia2006tensor}
\bibinfo{author}{Geremia, J.}, \bibinfo{author}{Stockton, J.~K.} \& \bibinfo{author}{Mabuchi, H.}
\newblock Tensor polarizability and dispersive quantum measurement of multilevel atoms.
\newblock \emph{\bibinfo{journal}{Phys. Rev. A}} \textbf{\bibinfo{volume}{73}}, \bibinfo{pages}{042112} (\bibinfo{year}{2006}).

\bibitem{borregaard2016scalable}
\bibinfo{author}{Borregaard, J.} \emph{et~al.}
\newblock Scalable photonic network architecture based on motional averaging in room temperature gas.
\newblock \emph{\bibinfo{journal}{Nat. Commun.}} \textbf{\bibinfo{volume}{7}}, \bibinfo{pages}{11356} (\bibinfo{year}{2016}).

\bibitem{Shaham2020}
\bibinfo{author}{Shaham, R.}, \bibinfo{author}{Katz, O.} \& \bibinfo{author}{Firstenberg, O.}
\newblock Quantum dynamics of collective spin states in a thermal gas.
\newblock \emph{\bibinfo{journal}{Phys. Rev. A}} \textbf{\bibinfo{volume}{102}}, \bibinfo{pages}{012822} (\bibinfo{year}{2020}).
\newblock \urlprefix\url{https://link.aps.org/doi/10.1103/PhysRevA.102.012822}.

\bibitem{Zeuthen2019}
\bibinfo{author}{Zeuthen, E.}, \bibinfo{author}{Polzik, E.~S.} \& \bibinfo{author}{Khalili, F.~Y.}
\newblock Gravitational wave detection beyond the standard quantum limit using a negative-mass spin system and virtual rigidity.
\newblock \emph{\bibinfo{journal}{Phys. Rev. D}} \textbf{\bibinfo{volume}{100}}, \bibinfo{pages}{062004} (\bibinfo{year}{2019}).
\newblock \urlprefix\url{https://link.aps.org/doi/10.1103/PhysRevD.100.062004}.

\bibitem{buchmann2016complex}
\bibinfo{author}{Buchmann, L.}, \bibinfo{author}{Schreppler, S.}, \bibinfo{author}{Kohler, J.}, \bibinfo{author}{Spethmann, N.} \& \bibinfo{author}{Stamper-Kurn, D.}
\newblock Complex squeezing and force measurement beyond the standard quantum limit.
\newblock \emph{\bibinfo{journal}{Phys. Rev. Lett.}} \textbf{\bibinfo{volume}{117}}, \bibinfo{pages}{030801} (\bibinfo{year}{2016}).

\bibitem{Alkis2023thesis}
\bibinfo{author}{Zoumis, A.}
\newblock \emph{\bibinfo{title}{Laser Noise Stabilisation for Low Frequency Quantum Back Action Cancellation}}.
\newblock \bibinfo{type}{Master's thesis}, \bibinfo{school}{University of Copenhagen} (\bibinfo{year}{2023}).

\bibitem{novikov2025hybrid}
\bibinfo{author}{Novikov, V.} \emph{et~al.}
\newblock Hybrid quantum network for sensing in the acoustic frequency range.
\newblock \emph{\bibinfo{journal}{Nature}} \textbf{\bibinfo{volume}{643}}, \bibinfo{pages}{955--960} (\bibinfo{year}{2025}).

\bibitem{li2018serf}
\bibinfo{author}{Li, J.} \emph{et~al.}
\newblock SERF atomic magnetometer--recent advances and applications: A review.
\newblock \emph{\bibinfo{journal}{IEEE Sensors Journal}} \textbf{\bibinfo{volume}{18}}, \bibinfo{pages}{8198--8207} (\bibinfo{year}{2018}).

\bibitem{jin2026four}
\bibinfo{author}{Jin, X.} \emph{et~al.}
\newblock Four Generations of Quantum Biomedical Sensors.
\newblock \emph{\bibinfo{journal}{arXiv preprint arXiv:2603.29944}}  (\bibinfo{year}{2026}).

\bibitem{9102993}
\bibinfo{author}{Zuo, S.} \emph{et~al.}
\newblock Ultrasensitive Magnetoelectric Sensing System for Pico-Tesla MagnetoMyoGraphy.
\newblock \emph{\bibinfo{journal}{IEEE Trans. Biomed. Circuits Syst.}} \textbf{\bibinfo{volume}{14}}, \bibinfo{pages}{971--984} (\bibinfo{year}{2020}).

\bibitem{Brasil2022}
\bibinfo{author}{Brasil, T.~B.}, \bibinfo{author}{Novikov, V.}, \bibinfo{author}{Kerdoncuff, H.} \emph{et~al.}
\newblock Two-colour high-purity Einstein-Podolsky-Rosen photonic state.
\newblock \emph{\bibinfo{journal}{Nat. Commun.}} \textbf{\bibinfo{volume}{13}} (\bibinfo{year}{2022}).
\newblock \urlprefix\url{http://dx.doi.org/10.1038/s41467-022-32495-7}.

\bibitem{Jia2023}
\bibinfo{author}{Jia, J.}, \bibinfo{author}{Novikov, V.}, \bibinfo{author}{Brasil, T.~B.} \emph{et~al.}
\newblock Acoustic frequency atomic spin oscillator in the quantum regime.
\newblock \emph{\bibinfo{journal}{Nat. Commun.}} \textbf{\bibinfo{volume}{14}} (\bibinfo{year}{2023}).
\newblock \urlprefix\url{http://dx.doi.org/10.1038/s41467-023-42059-y}.

\bibitem{grimaldi2025coherent}
\bibinfo{author}{Grimaldi, A.}, \bibinfo{author}{Novikov, V.}, \bibinfo{author}{Brasil, T.~B.} \& \bibinfo{author}{Polzik, E.~S.}
\newblock Coherent phase control of two-color continuous variable entangled light.
\newblock \emph{\bibinfo{journal}{arXiv preprint arXiv:2508.03303}}  (\bibinfo{year}{2025}).

\bibitem{jiathesis}
\bibinfo{author}{Jia, J.}
\newblock \emph{\bibinfo{title}{Conditional broadband quantum noise reduction with negative mass spin oscillators}}.
\newblock \bibinfo{type}{Ph.d. thesis}, \bibinfo{school}{Copenhagen University} (\bibinfo{year}{2024}).
\newblock \urlprefix\url{https://nbi.ku.dk/english/theses/phd-theses/jun-jia/}.

\bibitem{yu2020quantum}
\bibinfo{author}{Yu, H.} \emph{et~al.}
\newblock Quantum correlations between light and the kilogram-mass mirrors of LIGO.
\newblock \emph{\bibinfo{journal}{Nature}} \textbf{\bibinfo{volume}{583}}, \bibinfo{pages}{43--47} (\bibinfo{year}{2020}).

\bibitem{Kimble2001}
\bibinfo{author}{Kimble, H.~J.}, \bibinfo{author}{Levin, Y.}, \bibinfo{author}{Matsko, A.~B.} \emph{et~al.}
\newblock Conversion of conventional gravitational-wave interferometers into quantum nondemolition interferometers by modifying their input and/or output optics.
\newblock \emph{\bibinfo{journal}{Phys. Rev. D}} \textbf{\bibinfo{volume}{65}}, \bibinfo{pages}{022002} (\bibinfo{year}{2001}).
\newblock \urlprefix\url{https://link.aps.org/doi/10.1103/PhysRevD.65.022002}.

\bibitem{gould2021optimal}
\bibinfo{author}{Gould, D.~W.} \emph{et~al.}
\newblock Optimal quantum noise cancellation with an entangled witness channel.
\newblock \emph{\bibinfo{journal}{Phys. Rev. Research}} \textbf{\bibinfo{volume}{3}}, \bibinfo{pages}{043079} (\bibinfo{year}{2021}).

\bibitem{solomon1991psd}
\bibinfo{author}{Solomon~Jr, O.~M.}
\newblock PSD computations using Welch’s method.
\newblock \emph{\bibinfo{journal}{NASA STI/Recon Technical Report N}} \textbf{\bibinfo{volume}{92}}, \bibinfo{pages}{10--2172} (\bibinfo{year}{1991}).

\end{thebibliography}

\section*{Acknowledgments}
This work has been supported by VILLUM FONDEN under a Villum Investigator Grant, grant no.\ 25880, by the Novo Nordisk Foundation through Copenhagen Center for Biomedical Quantum Sensing, grant number NNF24SA0088433 and through ‘Quantum for Life’ Center, grant NNF20OC0059939. We acknowledge Valeriy Novikov and Sergey Fedorov for insightful discussions. 

\section*{Author Contributions}
J.J., T.B.B., M.Bo., A.G., and L.M. built the experimental setup and performed the measurements and data acquisition. J.J. led the work on atomic spin-oscillator magnetometer and its calibration; T.B.B. led the work on EPR-entangled light source. J.J. performed the data analysis. J.J. and E.Z. developed the analytical model. M.Ba. fabricated the cesium vapour cell. J.H.M. contributed valuable discussions. E.S.P. conceived and supervised the project. J.J., E.Z., and E.S.P. wrote the manuscript with contributions from all authors. All authors discussed the results and commented on the manuscript.


\section*{Author Information}
The authors declare no competing financial interests. Correspondence and requests for materials should be addressed to polzik@nbi.ku.dk.

\setcounter{figure}{0}
\renewcommand\theHfigure{sec.\thefigure}
\renewcommand{\figurename}{Extended Data Fig.}
\renewcommand{\tablename}{Extended Data Table}

\begin{figure*}[h] 
    \centering
    \includegraphics[width=0.8\linewidth]{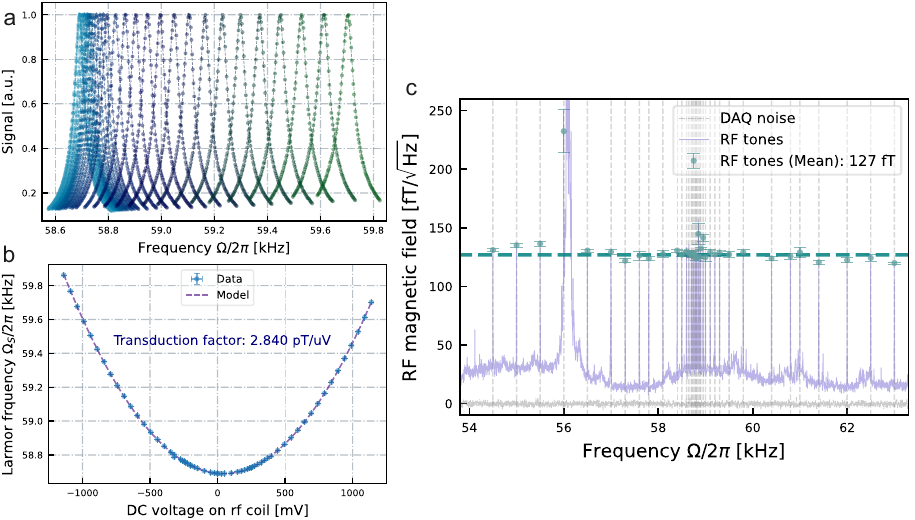}
    \caption{\textbf{Calibration of multi-radio-frequency (RF) magnetic tones.} \textbf{a,} Atomic MORS signals as functions of the DC voltage applied to the RF calibration coil. \textbf{b}, Larmor frequencies extracted from each MORS signal are plotted as functions of applied bias voltage. The extracted values are fitted with a parabolic function, from which the transduction factor-from DC voltage to bias magnetic field, is determined. \textbf{c,} The applied RF magnetic fields shown as the purple trace, indicated by the vertical dashed lines, are analyzed using the Welch spectrum. The RF magnetic amplitudes are calculated from the amplitude spectral densities and the calibrated field transduction factor. Electronic noise (shown as gray trace) from the data acquisition card is removed during the calibration. The median values with error bars at each RF frequency, shown as teal circles in panel~\textbf{c}, are calculated from seven independent measured traces. The mean RF field is determined by fitting the data with a constant offset (dashed horizontal teal line). Each trace is recorded for 50 seconds.}
    \label{fig:rf tone claibrations}
\end{figure*}

\begin{figure*}
    \centering
    \includegraphics[width=0.8\linewidth]{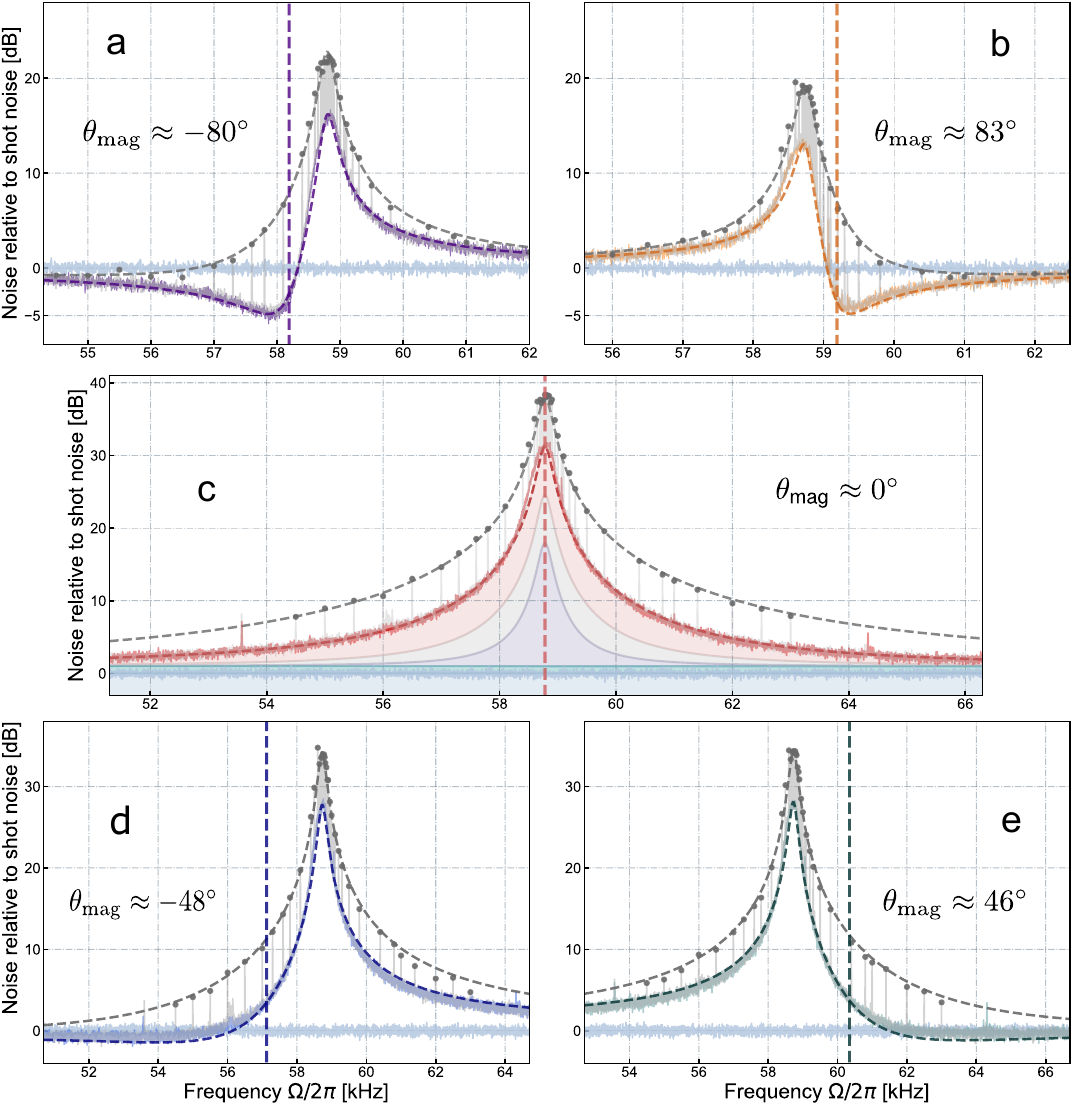}
    \caption{\textbf{Calibration of spin transfer function. } Spin noise spectra measured at different atomic detection phases (a, b, c, d, e) are shown, together with atomic dynamics driven by multi-tone RF magnetic fields of equal tone amplitude (gray dots in each subplot). The atomic response function to the RF tones is calibrated by fitting the spin noise both with (grey dashed curves) and without the RF field excitation (shown by colored dashed curves). The vertical dashed line in each panel marks the frequency at which each configuration reaches its minimum noise-equivalent sensitivity. The colour legend for the various noise contributions is the same as in main-text Fig.~\ref{fig:quantum enhancement with variational readout}.}
    \label{fig:Extended_fig_2}
\end{figure*}

\begin{figure*}
    \centering
    \includegraphics[width=0.9\linewidth]{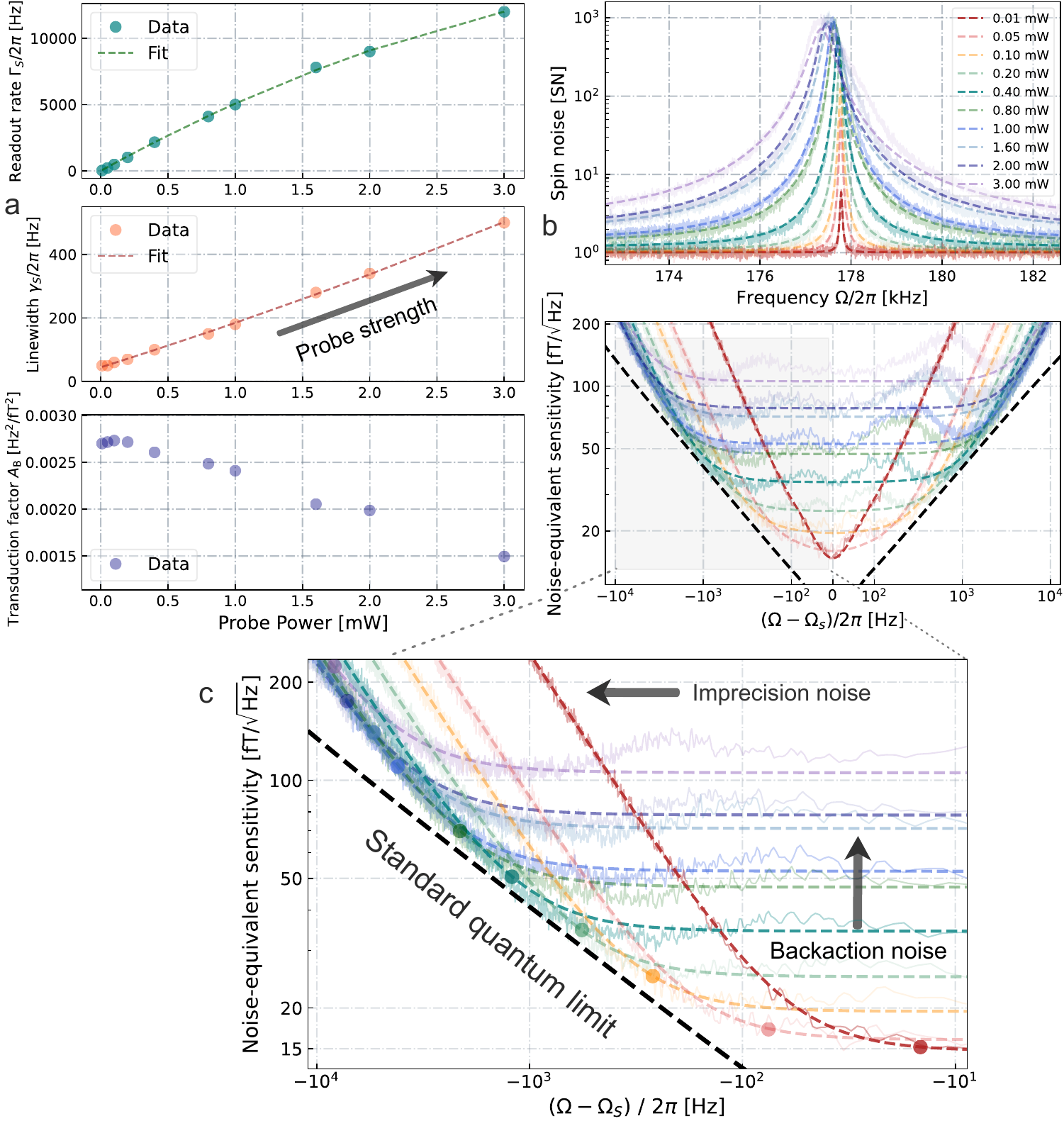}    \caption{\textbf{Reconstruction and validation of the standard quantum limit (SQL)}. Characterization of the optical magnetometer's performance across different probe strengths, and validation of the SQL from independently calibrated system parameters.  \textbf{a.} Extraction of the atomic readout rate $\Gamma_{S}$, total decoherence rate $\gamma_{S}$, and atomic transduction factor $A_{B}$ from the spin-noise spectrum, with and without multi-tone RF excitations, as a function of probe power. \textbf{b.} Measured spin-noise spectra in shot-noise units (solid) and corresponding theoretical predictions (dashed). The lower panel displays the magnetic-field sensitivity reconstructed from these fitted parameters. \textbf{c.} Probe-power dependence of the reconstructed magnetic-field sensitivity. Colored circles indicate the frequency ranges where the sensitivity at that specific probe power outperforms the other configurations. The predicted SQL (dashed black line) is calculated from independently calibrated system parameters using the dark decoherence rate and transduction factor calibrated at 1\,mW probe. The overall detection efficiency extracted from fits is $\eta^{\mathrm out}_{\mathrm{mag}}=$ 0.9, and the extracted thermal occupation number is $n_{S}= 1.8$.} 
    \label{fig:SQL validation}
\end{figure*}

\begin{figure*}
    \centering
    \includegraphics[width=0.8\linewidth]{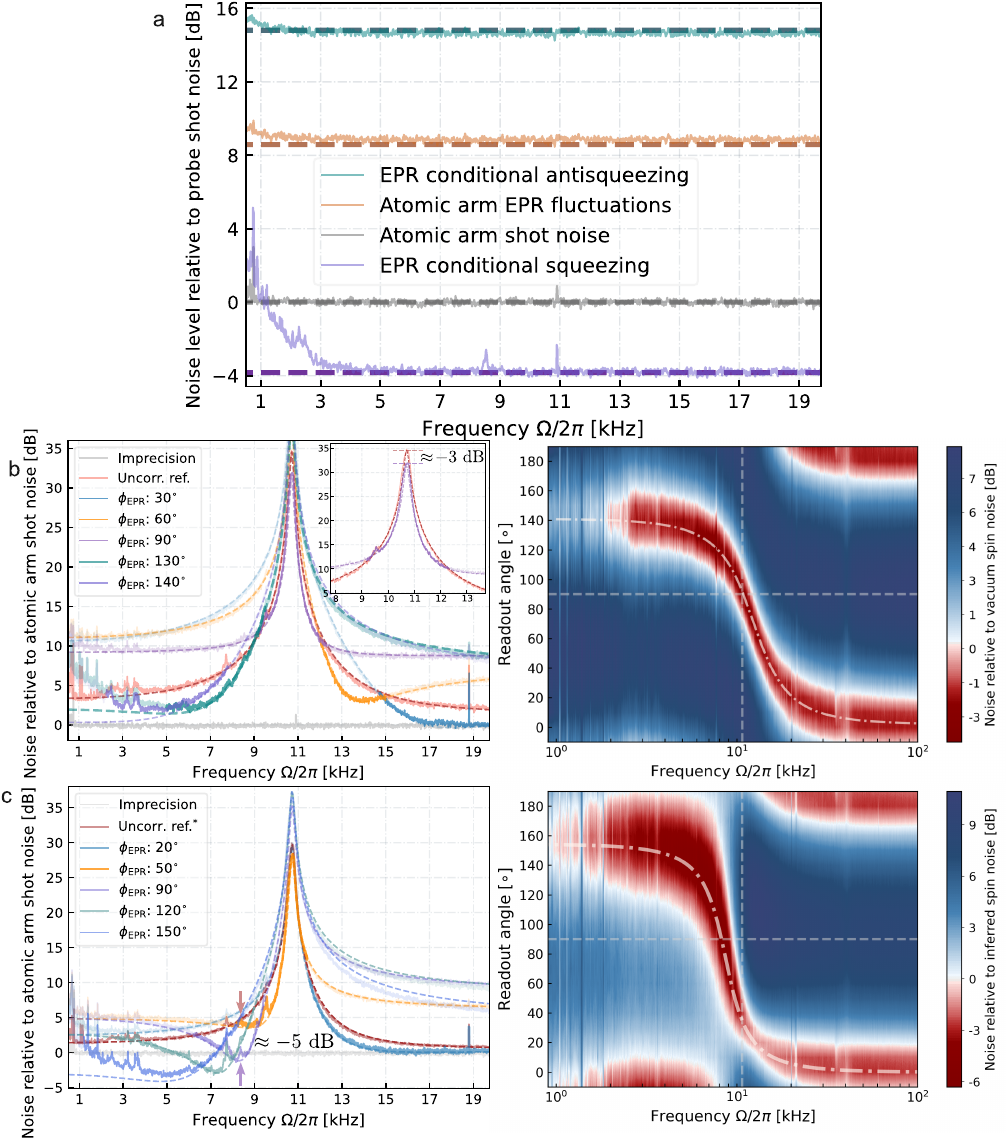}
    \caption{\textbf{Conditional spin noise spectra with EPR inference and hybrid correlation engineering.} \textbf{a}, Performance of the EPR-entangled light source with the spin oscillator detuned to 1 MHz. The grey trace shows probe fluctuations driven by vacuum noise. The orange trace shows the same probe channel under EPR injection. The purple trace indicates the conditional quantum noise obtained by quantum inference using the correlations with the parallel 1064nm arm. The teal trace shows anti-squeezed noise obtained when the two channels are combined oppositely. \textbf{b,} Conditional spin-noise spectra for different homodyne detection phases in the parallel EPR channel (colored curves), compared with the uncorrelated spin readout reference (red) [Uncorr.ref.]. Theory predictions are shown by dashed curves. Inset: zoom around the Larmor resonance showing the minimum conditional noise (-3 dB) using quantum correlation at $\phi_{\text{EPR}} = 90^{\circ}$. All spectra are plotted in shot noise units (dB scale). Contour plot of conditional noise relative to the vacuum-driven spin noise, plotted as a function of Fourier frequency and EPR homodyne readout angle. Red regions indicate noise level below the vacuum-probed reference. \textbf{c,} Hybrid configuration combining variational readout and EPR conditioning. Vacuum-driven spin-noise reference [Uncorr.ref.$^{*}$] is rescaled to account for signal reduction at the intermediate angle. Dashed curves show theoretical predictions. Contour plot of the conditional squeezing relative to this rescaled spin-noise reference. Red regions indicate the conditional squeezing, with darker shading corresponding to stronger squeezing level. More than 5 dB squeezing is observed about $\approx3$ kHz below the Larmor frequency, and the squeezing band extends down to $\sim1$ kHz.} 
    \label{fig:extend_figure_4}
\end{figure*}

\begin{figure*}
    \centering
    \includegraphics[width=0.9\linewidth]{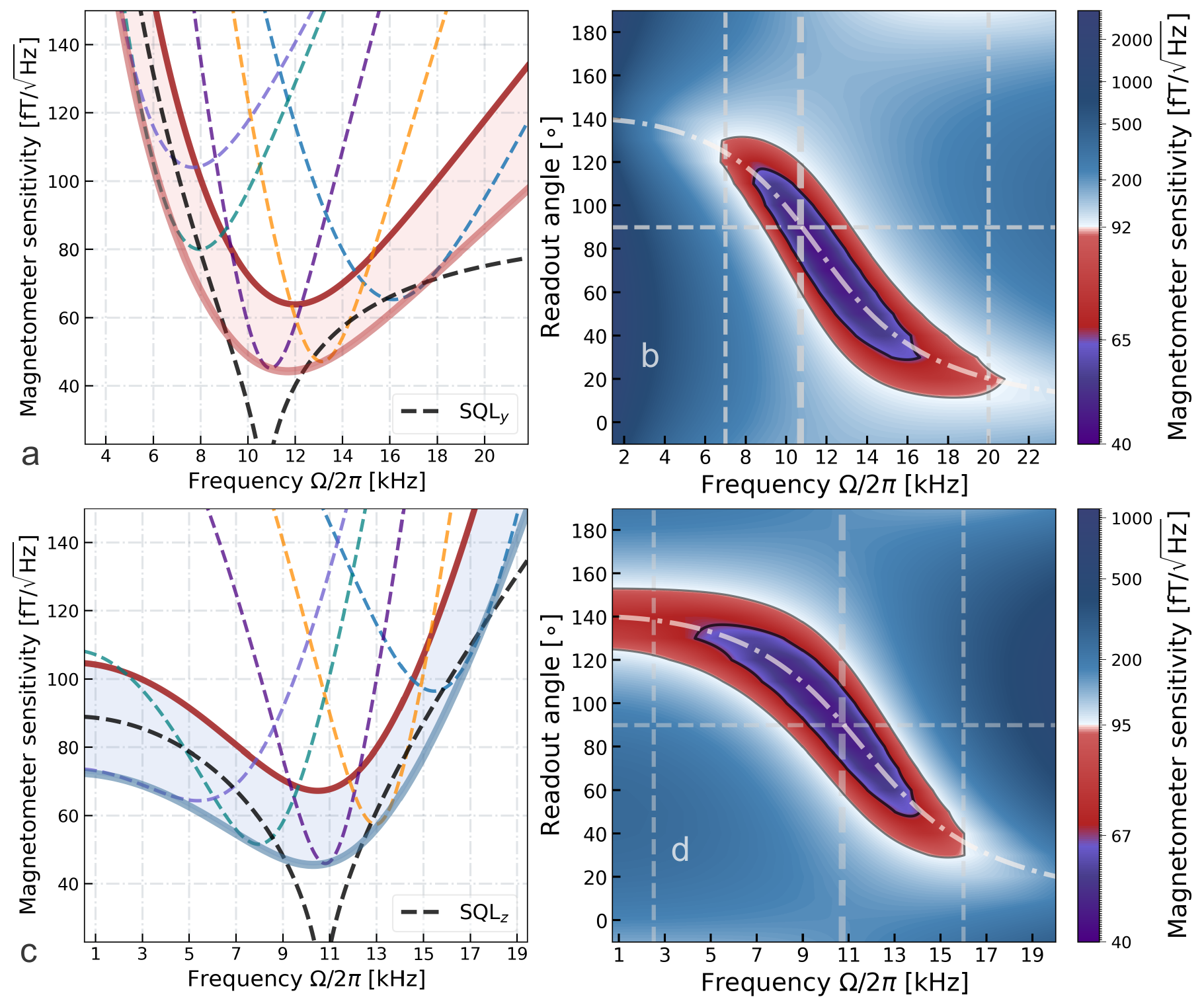}
    \caption{\textbf{Predicted vector magnetometer sensitivity.} \textbf{a,c} Reconstructed optical magnetometer sensitivity for signal along the transverse spin components $y$ (a) and $z$ (c). The dark solid curve denotes the uncorrelated spin noise reference, while dashed curves show the conditioned sensitivity for different EPR conditioning phases. The shaped regions indicate sensitivity enhancement obtained with the optimal conditioning phase. \textbf{b,d} Contour plots showing the theoretical predicted sensitivity as a function of EPR readout angles and sideband frequency for the $y$ (b) and $z$ (d) components.  }
    \label{fig:Extended_Figure_5}
\end{figure*}
\FloatBarrier

\begin{table*}[t]
    \centering
    \begin{tabular}{lll}
        \toprule
        \RR{8.5cm}{\textbf{Parameter}} & \RR{3cm}{\textbf{Symbol}} & \RL{3cm}{\textbf{Value}} \\
        \midrule
        \multicolumn{3}{c}{\textbf{Hybrid system \& detection}} \\
        \midrule
        \RR{8.5cm}{Two-mode squeezing factor} & \RR{3cm}{$r$} & \RL{3cm}{1.43} \\
        \RR{8.5cm}{Propagation efficiency before atoms} & \RR{3cm}{$\eta^{\mathrm{in}}_{\mathrm{mag}}$} & \RL{3cm}{0.89} \\
        \RR{8.5cm}{Overall efficiency after atoms} & \RR{3cm}{$\eta^{\mathrm{out}}_{\mathrm{mag}}$} & \RL{3cm}{0.90} \\
        \RR{8.5cm}{Parallel EPR channel efficiency (propagation \& detection)} & \RR{3cm}{$\eta_{\mathrm{EPR}}$} & \RL{3cm}{0.92} \\
        \RR{8.5cm}{Atomic channel efficiency (propagation \& detection)} & \RR{3cm}{$\eta_{\mathrm{mag}}\,(=\eta^{\mathrm{out}}_{\mathrm{mag}}\eta^{\mathrm{in}}_{\mathrm{mag}})$} & \RL{3cm}{0.8} \\
        \RR{8.5cm}{QWP phase} & \RR{3cm}{$\theta_{\mathrm{mag}}$} & \RL{3cm}{$0^{\circ}$, $-55^{\circ}$} \\
        \RR{8.5cm}{Signal LO power} & \RR{3cm}{} & \RL{3cm}{\SI{1}{\milli\watt}} \\
        \RR{8.5cm}{Idler LO power} & \RR{3cm}{} & \RL{3cm}{\SI{1}{\milli\watt}} \\
        \midrule
        \multicolumn{3}{c}{\textbf{Atomic spin oscillator}} \\
        \midrule
        \RR{8.5cm}{Larmor frequency} & \RR{3cm}{$\Omega_\mathrm{S}/2\pi$ [kHz]} & \RL{3cm}{10.7, 58.8} \\
        \RR{8.5cm}{Spin readout rate} & \RR{3cm}{$\Gamma_\mathrm{S}/2\pi$ [kHz]} & \RL{3cm}{9.3, 8.2} \\
        \RR{8.5cm}{Spin decoherence rate} & \RR{3cm}{$\gamma_\mathrm{S}/2\pi$ [kHz]} & \RL{3cm}{0.19, 0.24} \\
        \RR{8.5cm}{Effective spin thermal occupation} & \RR{3cm}{$n_\mathrm{S}$} & \RL{3cm}{3.4, 1.8} \\
        \RR{8.5cm}{Spin broadband readout rate} & \RR{3cm}{$\Gamma_\mathrm{bb}/2\pi$ [kHz]} & \RL{3cm}{2, 3.5} \\
        \RR{8.5cm}{Spin broadband decoherence rate} & \RR{3cm}{$\gamma_\mathrm{bb}/2\pi$ [kHz]} & \RL{3cm}{145} \\
        \RR{8.5cm}{Effective spin broadband occupation} & \RR{3cm}{$n_\mathrm{bb}$} & \RL{3cm}{3.4, 1.8} \\
        \RR{8.5cm}{Transduction factor} & \RR{3cm}{$A_{\mathrm{B}}$ [$\times 10^{-3}\,\mathrm{Hz}^{2}/\mathrm{fT}^{2}$]} & \RL{3cm}{2.6(5), 2.3(4)} \\
        \RR{8.5cm}{Probe field detuning} & \RR{3cm}{$\Delta_{\mathrm{mag}}/2\pi$} & \RL{3cm}{\SI{1.6}{\giga\hertz}} \\
        \RR{8.5cm}{Magnetic transduction coefficient} & \RR{3cm}{$\alpha_\mathrm{B}$ [$\mathrm{pT}/\upmu\mathrm{V}$]} & \RL{3cm}{2.84} \\
        \RR{8.5cm}{Probe input polarization} & \RR{3cm}{$\beta$} & \RL{3cm}{$45^{\circ}$} \\
        \RR{8.5cm}{Spin polarization} & \RR{3cm}{} & \RL{3cm}{$\sim$82\%} \\
        \bottomrule
    \end{tabular}
    \caption{Summary of notations and experimental parameters for the quantum enhanced magnetometer.}
    \label{tab:hybrid quantum sensing}
\end{table*}

\end{document}


\title{Supplementary Material for\\``Entanglement-enhanced optical magnetometry beyond the standard quantum limit''}
\author{Jun Jia}
\thanks{These authors contributed equally}
\affiliation{Niels Bohr Institute, University of Copenhagen, Blegdamsvej 17, DK-2100 Copenhagen Ø, Denmark}
\author{Túlio Brito Brasil}
\thanks{These authors contributed equally}
\affiliation{Niels Bohr Institute, University of Copenhagen, Blegdamsvej 17, DK-2100 Copenhagen Ø, Denmark}
\author{Maimouna Bocoum}
\altaffiliation{Present address: Institut Langevin, ESPCI Paris, Université PSL, Sorbonne Université, Université Paris Cité, CNRS, 75005 Paris, France.}
\affiliation{Niels Bohr Institute, University of Copenhagen, Blegdamsvej 17, DK-2100 Copenhagen Ø, Denmark}
\author{Andrea Grimaldi} 
\altaffiliation{Present address: INFN, Sezione di Padova, I-35131 Padova, Italy.}
\affiliation{Niels Bohr Institute, University of Copenhagen, Blegdamsvej 17, DK-2100 Copenhagen Ø, Denmark}
\author{Laurits Møberg} 
\affiliation{Niels Bohr Institute, University of Copenhagen, Blegdamsvej 17, DK-2100 Copenhagen Ø, Denmark}
\author{Mikhail Balabas}
\affiliation{Niels Bohr Institute, University of Copenhagen, Blegdamsvej 17, DK-2100 Copenhagen Ø, Denmark}
\author{J\"{o}rg Helge M\"{u}ller}
\affiliation{Niels Bohr Institute, University of Copenhagen, Blegdamsvej 17, DK-2100 Copenhagen Ø, Denmark}
\author{Emil Zeuthen}
\affiliation{Niels Bohr Institute, University of Copenhagen, Blegdamsvej 17, DK-2100 Copenhagen Ø, Denmark}
\author{Eugene Simon Polzik}
\affiliation{Niels Bohr Institute, University of Copenhagen, Blegdamsvej 17, DK-2100 Copenhagen Ø, Denmark}

\begin{abstract}
This document provides a detailed theoretical model describing sub-SQL operation of an optical magnetometer enabled by engineered quantum correlations.
\end{abstract}

\maketitle
\tableofcontents
\section{Linear spin oscillator model}
Magnetic field sensing in this work is achieved via continuous Faraday probing of an optically polarized spin ensemble. A linearly polarized probe beam propagating along the z-axis experiences circular birefringence induced by collective spin projections $F_{z}$, resulting in a polarization rotation that directly encodes the spin signal. In the presence of a bias magnetic field $B_{\mathrm{0}}$ applied along the x-axis, the collective spin precesses in the $y$-$z$ plane at the Larmor frequency $\Omega_{S} = g_{\mathrm{B}} B_{\mathrm{0}}$, such that the time-dependent spin projection $F_{z}$ is continuously mapped onto the probe via the Faraday interaction. 
In addition, weak transverse radio-frequency (RF) magnetic fields applied along the $y$-axis couple to the spin oscillator and drive coherent oscillations of $F_{z}$ near the Larmor resonance, providing controlled input signals for calibration and characterization of the magnetometer response.
In the experiment, multi-tone RF excitations are used to calibrate the transfer function, whereas in the following theoretical analysis, a single-tone RF drive is considered for simplicity.
\subsection{Bloch equation and linearization}
The collective spin dynamics under magnetic driving  ($\sim g_{\mathrm{B}}B_{\mathrm{RF}}\mathrm{cos}(\Omega_{\mathrm{RF} }t) \hat{y}\cdot \vec{F}$) and optical probing ($\sim g_{S}S_{z}  F_{z}$) are described by the Bloch equation
\begin{equation}
    \frac{d \vec{F}}{dt} = -g_{B} (B_{0}\hat{x}+B_{\mathrm{RF}}\mathrm{cos}(\Omega_{\mathrm{RF}}t)\hat{y})\times  \vec{F}-g_{S}S_{z}\hat{z}\times  \vec{F} - \frac{\gamma_{S}}{2}(F_{y}\hat{y}+F_{z}\hat{z}) -\Gamma_{\mathrm{op}}(F_{x}-F_{x}^{ss})\hat{x}+ \sqrt{\gamma_{S}}(f_{y}\hat{y}+f_{z}\hat{z}),
\end{equation}
where $g_{\mathrm{B}}$ is the gyromagnetic ratio \cite{steck2003cesium}, $g_{S}$ is the light-atom coupling constant \cite{Thomas2021,novikov2025hybrid}, $\gamma_{S}$ is the transverse decoherence rate (including dark decay rate, probe power induced spontaneous emission, imperfect motional averaging in the presence of magnetic-field inhomogeneity), and $\Gamma_{\mathrm{op}}$ describes optical pumping towards steady state polarization $F_{x}^{ss}$. The operators $f_{y}$ and $f_{z}$ represent the Langevin force associated with the spin decoherence.
The second term describes the quantum backaction arising from the AC-Stark shift induced by probe fluctuations.
In the regime of highly polarized spin ensemble ($F_y,F_z \ll F_{x}^{ss}$), linearly polarized probe ($S_{y}, S_{z}\ll S_{x}$), and weak RF drive ($B_{\mathrm{RF}}\ll B_{0}$), the dynamics can be linearized around the stationary spin state $F_{x} \approx  F_x^{ss}$. Neglecting higher-order terms \cite{Julsgaard2003thesis}, the transverse spin components obey the dynamics
 \begin{align}
        \frac{dF_{y}}{dt} &= -g_{B}B_{0}F_{z} -g_{S}S_{z}F_{x}^{ss}-\frac{\gamma_{S}}{2}F_{y}+\sqrt{\gamma_{S}}f_{y},\\
        \frac{dF_{z}}{dt} &= g_{B}B_{0}F_{y} -g_{B}B_{\mathrm{RF}}F_{x}^{ss}\cos(\Omega_{\mathrm{RF}} t)-\frac{\gamma_{S}}{2}F_{z}+\sqrt{\gamma_{S}}f_{z}.
\end{align}
These equations describe an RF force driven, damped harmonic oscillator formed by two transverse spin components. Taking the Fourier transform ($d/dt \rightarrow -i\Omega$), the spin response $F_{z}(\Omega)$ can be written in terms of two susceptibilities:
\begin{equation}
    F_{z}(\Omega) = -g_{B}F_{x}^{ss}\rho_{S}(\Omega) B_{\mathrm{RF}}(\Omega) - g_{S}F_{x}^{ss}\chi_{S}(\Omega)S_{z} (\Omega) +\sqrt{\gamma_{S}}(\rho_{S}(\Omega)f_{z}(\Omega)+\chi_{S}(\Omega)f_{y}(\Omega)),
\end{equation}
here $B_{\mathrm{RF}}(\Omega)$ denotes the externally applied classical magnetic field used for signal injection and calibration, and the susceptibilities $\rho_{S}(\Omega)$ and $\chi_{S}(\Omega)$ are:
\begin{align}
    \rho_{s}(\Omega) &= \frac{\gamma_{S}/2 - i\Omega}{\Omega^{2}_{S} + (\gamma_{S}/2)^2 -\Omega^{2} -i \gamma^{}_{S}\Omega},\\
     \chi_{s}(\Omega) &= \frac{\Omega_{S}}{\Omega^{2}_{S} + (\gamma_{S}/2)^2 -\Omega^{2} -i \gamma^{}_{S}\Omega}.
\end{align}
$\Omega_{S}$ is the atomic Larmor frequency defined by $g_{B}B_{0}$; the Faraday readout can be described as (neglecting retardation):
\begin{align}
    S_{y}^{\mathrm{out}}(\Omega) &= S_{y}^{\mathrm{in}}(\Omega) - g_{S}S_{x}F_{z} (\Omega),\\
    S_{z}^{\mathrm{out}}(\Omega) &= S_{z}^{in}(\Omega).
\end{align}
To describe the system within a more convenient quantum measurement framework, we introduce dimensionless canonical quadratures for both the collective spin and probe field.
The spin transverse quadratures are defined as:
\begin{align}
    X_{S} = \frac{F_{z}}{\sqrt{F_{x}^{ss}}}, P_{S} =- \frac{F_{y}}{\sqrt{F_{x}^{ss}}},
\end{align}
with commutation relation $\comm{X_{S}}{P_{S}} =i\delta(\Omega+\Omega')$.
Similarly, the optical probe is described by the canonical quadratures
\begin{align}
    X_{L} = \frac{S_{z}}{\sqrt{S_{x}}}, P_{L} = -\frac{S_{y}}{\sqrt{S_{x }}},
\end{align}
which also satisfy $\comm{X_{L}}{P_{L}} = i\delta (\Omega+\Omega')$.
With this normalization, the Fourier-domain equation for the measured spin quadrature then reads as:
\begin{equation}
    X_{S}(\Omega) = -\sqrt{\Gamma_{S}} \chi_{S}(\Omega)X_{L}^{\mathrm{in}}(\Omega) - \sqrt{A_{B}}\rho_{S}(\Omega)B_{RF}(\Omega) + \sqrt{\gamma_{S}}(\rho_{S}(\Omega)f^{X_{S}}+\chi_{S}(\Omega)f^{P_{S}}),
\end{equation}
where $\Gamma_{S} = g_{S}^{2}S_{x}F_{x}^{ss}$ is the probe readout rate, $A_{B} = g_{B}^{2}F_{x}^{ss}$ represents the field to spin displacement transduction factor. $f^{X_{S}} = f_{z}/\sqrt{F_{x}^{ss}}$ and $f^{P_{S}}=-f_{y}/\sqrt{F_{x}^{ss}}$ are dimensionless Langevin force.\\ 
The optical input-ouput relations are
\begin{align}
    X_{L}^{\mathrm{out}}(\Omega) =& X_{L}^{\mathrm{in}}(\Omega),\\
    P_{L}^{\mathrm{out}}(\Omega) =& P_{L}^{\mathrm{in}}(\Omega) - \sqrt{\Gamma_{S} }X_{S}(\Omega) ,
\end{align}
substituting the spin response gives the following result
\begin{equation}
    P_{L}^{\mathrm{out}}(\Omega) =P_{L}^{\mathrm{in}}(\Omega) + \Gamma_{S} \chi_{S}(\Omega)X_{L}^{\mathrm{in}}(\Omega) + \sqrt{\Gamma_{S}A_{B}}\rho_{S}(\Omega)B_{RF}(\Omega) - \sqrt{\Gamma_{S}\gamma_{S}}(\rho_{S}(\Omega)f^{X_{S}}+\chi_{S}(\Omega)f^{P_{S}}).
\label{eq:inputoutput}
\end{equation}

\section{Power spectral density and noise-equivalent magnetic sensitivity}
Here we introduce the power spectral density (PSD) analysis of the optical magnetometer discussed in this work. We also specify the spectral properties of vacuum source, two-mode squeezed EPR input fields, and spin Langevin force noise.
The spectrum analysis is calculated using the symmetrized cross-spectral density $S_{A,B}$ \cite{Danilishin2012} of two operators $A$ and $B$ as
\begin{equation}
    2\pi S_{A,B}(\Omega)\delta(\Omega-\Omega')= \frac{1}{2}\left<A(\Omega)B^{\dagger}(\Omega')+B^{\dagger}(\Omega')A(\Omega)\right>,
\end{equation}  
with the Hermitian operators satisfying $A^{\dagger}(\Omega) = A(-\Omega)$. 
Using the introduced canonical normalization, both optical vacuum and spin ground state fluctuations have symmetrized spectral densities equal to $\frac{1}{2}$.
For a vacuum limited probe field, the input quadratures satisfy:
\begin{equation}
    S_{\mathrm{vac}}(\Omega)= S_{X^{\mathrm{in}}_{L}, X^{\mathrm{in}}_{L}}(\Omega) = S_{P^{\mathrm{in}}_{L}P^{\mathrm{in}}_{L}}(\Omega) = \frac{1}{2},  S_{X^{\mathrm{in}}_{L}, P^{\mathrm{in}}_{L}}(\Omega) = 0.
\end{equation}
When the probe's orthogonal polarization mode is replaced by one of an EPR entangled state, the spectra characterize both the fluctuations and correlations between the atomic sensing channel and optical conditioning channel \cite{Danilishin2019}:
\begin{align}
    S_{X_{L,\mathrm{mag}},X_{L,\mathrm{mag}}}(\Omega) &=  S_{P_{L,\mathrm{mag}} ,P_{L,\mathrm{mag}}}(\Omega)   =  S_{X_{L,\mathrm{EPR}},X_{L,\mathrm{EPR}}}(\Omega) = S_{P_{L,\mathrm{EPR}},P_{L,\mathrm{EPR}}}(\Omega)  = \frac{\mathrm{cosh(2r)}}{2},\\
    S_{X_{L,\mathrm{mag}},X_{L,\mathrm{EPR}}}(\Omega) &=  - S_{P_{L,\mathrm{mag}},P_{L,\mathrm{EPR}}}(\Omega) = \frac{\mathrm{sinh(2r)}}{2}.
\end{align}
Here, `mag' and `EPR' indicate the optical fields interacting with the atomic ensemble and conditioning measurement channels, respectively, and r is the squeezing parameter of a two-mode squeezed vacuum state.
The spin Langevin forces are modelled as Markovian noise sources with white spectra
\begin{align}
    S_{f_{s}^{X_{s}},f_{s}^{X_{s}}}(\Omega) &= S_{f_{s}^{P_{s}},f_{s}^{P_{s}}}(\Omega) = n_{s}+ \frac{1}{2},\\
    S_{f_{s}^{X_{s}}, f_{s}^{P_{s}}}(\Omega) &= 0,
\end{align}
where $n_{s}$ denotes the effective thermal occupation associated with the impurity of the coherent spin state. 
Using the previously derived input-output relations.(\ref{eq:inputoutput}), we calculate the noise spectrum of output field phase quadrature,
\begin{align}
S_{P_{L}^{\mathrm{out}}, P_{L}^{\mathrm{out}}}(\Omega) = &S_{\mathrm{imp}}(\Omega)+S_{\mathrm{qba}}(\Omega)+S_{\mathrm{sig}}(\Omega)+S_{\mathrm{spin}}(\Omega),\\
    S_{\mathrm{imp}}(\Omega) = &S_{P_{L}^{\mathrm{in}}, P_{L}^{\mathrm{in}}}(\Omega),\\
    S_{\mathrm{qba}}(\Omega) = &\Gamma_{S}^{2}\abs{\chi_{S}(\Omega)}^{2} S_{X_{L}^{\mathrm{in}}, X_{L}^{\mathrm{in}}}(\Omega),\\
    S_{\mathrm{sig}}(\Omega) =& \Gamma_{S}A_{B}\abs{\rho_{S}(\Omega)}^2S_{B_{\mathrm{RF}}}(\Omega),\\
    S_{\mathrm{spin}}(\Omega)=&\Gamma_{S}\gamma_{S}(\abs{\rho_{S}(\Omega)}^{2}+\abs{\chi_{S}(\Omega)}^{2}) (n_{S}+1/2) + \Gamma_{bb}\gamma_{bb}(\abs{\rho_{bb}(\Omega)}^{2}+\abs{\chi_{bb}(\Omega)}^{2}) (n_{bb}+1/2).
\end{align}
Note that, besides the decoherence-associated spin noise, there exists an additional broadband decoherence rate $\gamma_{bb}$ arising from the inhomogeneous interaction between the probe light and moving atoms confined in the vapour cell, which introduces a broadband decoherence channel beyond the intrinsic spin dynamic. Its decoherence rate is defined as $\gamma_{bb}$, with an interaction readout rate of $\Gamma_{bb}$. $\chi_{bb}(\Omega)$ and $\rho_{bb}(\Omega)$ are faster decoherence rate associated susceptibilities.
From the calibrated total measurement noise, the noise equivalent magnetic-field sensitivity of the optical magnetometer along y-axis can be reconstructed as
\begin{equation}
    \delta B_{y}(\Omega) = \frac{\sqrt{S_{\mathrm{imp}}(\Omega)+S_{\mathrm{qba}}(\Omega)+S_{\mathrm{spin}}(\Omega)}}{\sqrt{\Gamma_{S}A_{B}}\abs{\rho_{S}(\Omega)}} .
\end{equation}
For completeness, we note that within the same linear-response framework for the orientation based vector optical magnetometer, an analogous noise-equivalent magnetic-field sensitivity can be defined for a transverse field applied along the $z$-axis. In this case, the field transduction is governed by a different spin susceptibility, reflecting the physics that the $B_{z}$ field couples to the orthogonal spin quadrature $F_{y}$ and reaches the measured observable via the Larmor precession. The corresponding sensitivity can therefore be written in an analogous form,
\begin{equation}
    \delta B_{z}(\Omega) = \frac{\sqrt{S_{\mathrm{imp}}(\Omega)+S_{\mathrm{qba}}(\Omega)+S_{\mathrm{spin}}(\Omega)}}{\sqrt{\Gamma_{S}A_{B}}\abs{\chi_{S}(\Omega)}} ,
\end{equation}
where $\chi_{S}(\Omega)$ denotes the susceptibility associated with this path of coupling. Comparing the two channels reveals an intrinsic anisotropy of the vector magnetometer frequency response, with the relative sensitivity set by $\abs{\rho_{S}(\Omega)/\chi_{S}(\Omega)} = \abs{(\frac{\gamma_{S}}{2}-i\Omega)}/\Omega_{S}$. This shows that the anisotropy is influenced by the competition between the analysis frequency and the Larmor frequency, and becomes increasingly pronounced away from the regime where these two are comparable as shown in Fig.~S1. Although the present experiment probes only the $y$-channel, this highlights a fundamental frequency-dependent anisotropy arising from distinct transduction coupling of spin transverse quadratures. This suggests that achieving optimal broadband sensitivity in a vector magnetometer may require multi-quadrature readout or use frequency-dependent measurement basis.
\section{Standard quantum limit}
When considering only the measurement-added quantum noise, the magnetic field sensitivity can be written as
\begin{equation}
    \delta B_{y} = \frac{\sqrt{S_{\mathrm{imp}}+S_{\mathrm{qba}}}}{\sqrt{\Gamma_{S}A_{B}}\abs{\rho_{S}(\Omega)}}  = \frac{\sqrt{1/2+\frac{\Gamma_{S}^{2}}{2} \abs{\chi_{S}(\Omega)}^2}}{\sqrt{\Gamma_{S}A_{B}}\abs{\rho_{S}(\Omega)}} \geq  \delta B_{y,\mathrm{SQL}} = \frac{\sqrt{\abs{\chi_{S}(\Omega)}}}{\sqrt{A_{B}}\abs{\rho_{S}(\Omega)}}.
\end{equation}
This expression defines the minimum achievable sensitivity, i.e., the standard quantum limit \cite{braginsky1967classical,braginsky1995quantum, khalili2021quantum}, in the absence of correlations between imprecision and backaction noise. Similarly, the SQL along the $z$-axis can be estimated as $\delta B_{z, \mathrm{SQL}} = 1/\sqrt{A_{B}\abs{\chi_{S}(\Omega)}} $
In the data analysis, the measured spin noise spectra, with and without RF excitation, are fitted in shot noise units. Because the experimentally calibrated spectra are expressed in shot-noise units (normalized to vacuum fluctuations of 1/2), the extracted transduction factor satisfies $A_{B,\mathrm{data}} = 2A_{B}$. Consequently, the SQL plotted in this work is given by $\delta B_{y,\mathrm{SQL}} = \sqrt{2\abs{\chi_{S}(\Omega)}/(A_{B,\mathrm{data}}\abs{\rho_{S}(\Omega)}^{2})}$. The sub-SQL performance analysis follows methods developed in gravitational-wave experiments~\cite{yu2020quantum}.
\section{Effect of losses and decoherence}
Optical losses are modeled as coupling to uncorrelated vacuum fluctuations via effective beam splitters with transmissivity $\eta$, such that
\begin{align}
X^{\mathrm{out}} &= \sqrt{\eta}\, X^{\mathrm{in}} + \sqrt{1-\eta}\, X_{\mathrm{vac}}, \\
P^{\mathrm{out}} &= \sqrt{\eta}\, P^{\mathrm{in}} + \sqrt{1-\eta}\, P_{\mathrm{vac}}.
\end{align}
The atomic channel is described by input propagation efficiency $\eta^{\mathrm{in}}_{\mathrm{mag}}$ and output efficiency $\eta^{\mathrm{out}}_{\mathrm{mag}}$, which includes optical losses and detector quantum efficiency. Input losses introduce vacuum fluctuations before interaction, contributing to both imprecision and additional quantum backaction, and degrading correlation with the EPR channel. The EPR conditioning arm is modeled by an overall efficiency $\eta_{\mathrm{EPR}}$.
These loss channels are incorporated into the full noise model used to fit the measured spectra and extract the magnetic-field transduction factor $A_{B}$, from which the SQL reference is constructed.
The SQL depends on the spin susceptibilities $\chi_{S}(\Omega)$ and $\rho_{S}(\Omega)$, which in general is modified by probe-induced power broadening as shown in Fig.~S2. In this work, we define the SQL using intrinsic (dark) decoherence rate, providing a consistent reference for the minimum achievable quantum noise across the analysed frequency range. At the off-resonant frequencies considered here, the effect of power broadening on the SQL is negligible.
In the ideal lossless case with only measurement-added noise, the sensitivity reaches the SQL at the optimal frequency. Intrinsic spin noise (projection and thermal) shifts the minimum sensitivity above the SQL, and optical losses further degrade the achievable sensitivity.
\begin{figure*}[htp!]
    \centering
    \includegraphics[width=0.80\linewidth]{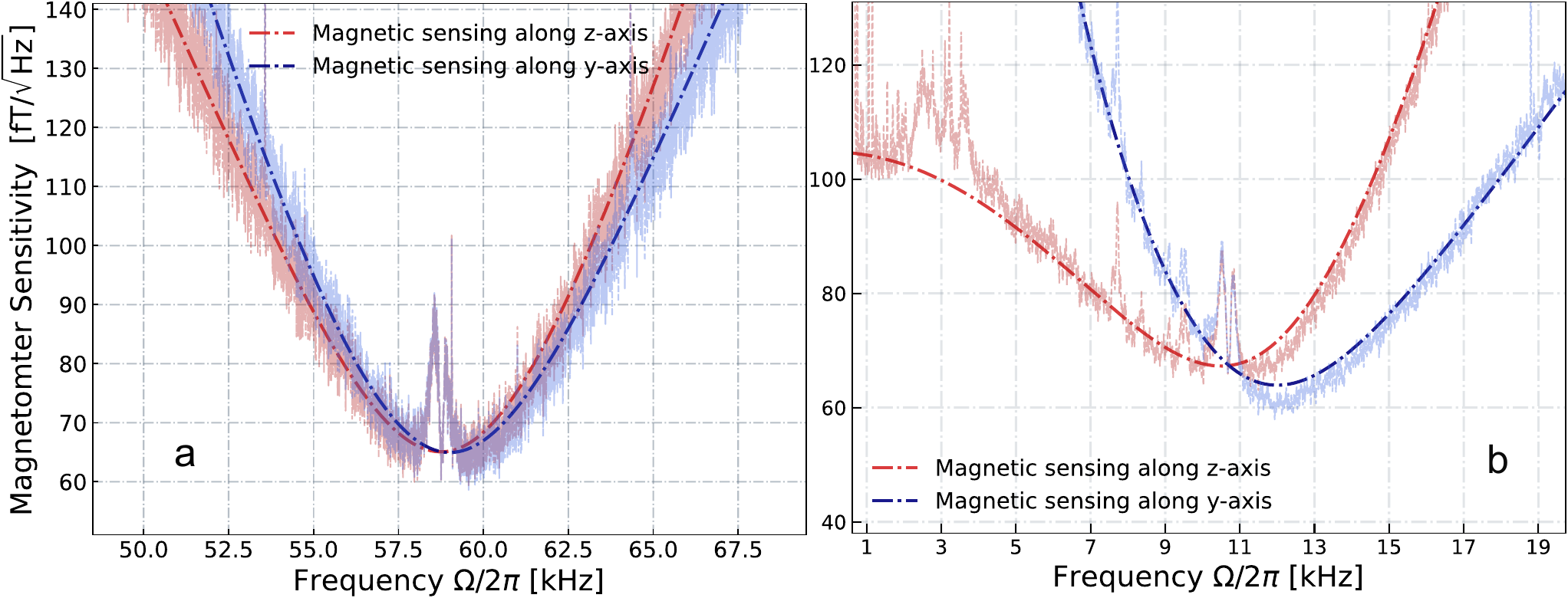}
    \caption{\textbf{Predicted vector optical magnetometer sensitivity at different Larmor frequencies.} \textbf{a \& b,} The optical magnetometer operates at Larmor frequencies of $\Omega_{S}/2\pi = 58\,\mathrm{kHz}$ and $10.7\,\mathrm{kHz}$, respectively.}
    \label{fig:vector magnetometer}
\end{figure*}

\begin{figure*}[htp!]
    \centering
    \includegraphics[width=0.60\linewidth]{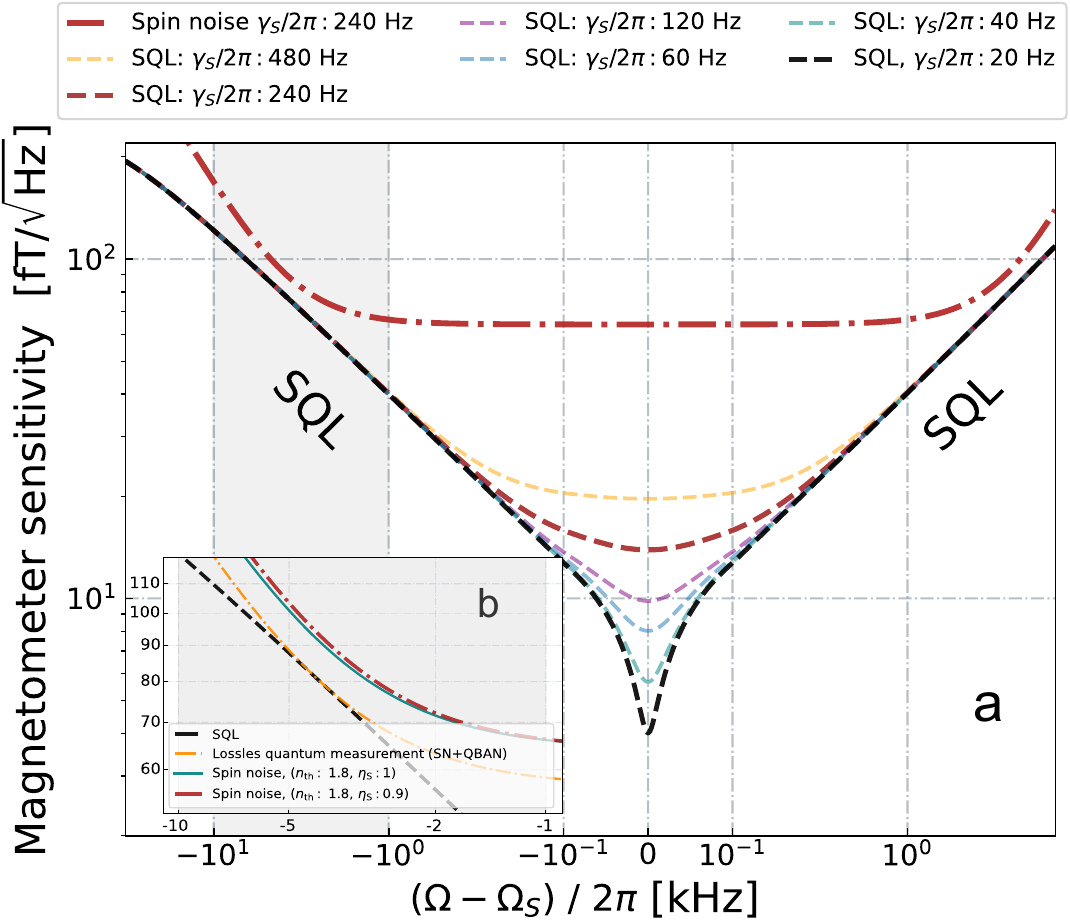}
    \caption{\textbf{SQL under different power broadening.} \textbf{a,} Predicted spin noise-equivalent magnetic field sensitivity and SQL for varying total decoherence under experimental conditions. \textbf{b,} Off-resonance sensitivity(-10 to 1 kHz). In the ideal lossless case ($\eta_{S}$), where only quantum backaction and imprecision noise are present, the measurement noise reaches the SQL at the optimal frequency. Including intrinsic spin noise (green curve) raises the minimum noise above the SQL, while finite detection efficiency (red dash–dotted curve) increases it further. Parameters are extrapolated from measurement at 1 mW probe.}
    \label{fig:SQL with difference power broadening}
\end{figure*}
\section{General measurement model}
All quantum-noise engineered configurations considered below share the same spin response functions, loss model, and noise decomposition introduced in the above sections. The differences between the variational-readout, EPR-conditioning, and hybrid configurations arise only from the choice of detected quadratures and from the presence or absence of inter/cross-channel quantum correlations.
For all configurations discussed below, the measured spectrum $S_{\mathrm{meas}}(\Omega)$, excluding the driven signal contribution $S_{\mathrm{sig}}(\Omega)$, can be expressed in terms of a common set of noise contributions,
\begin{equation}
    S_{\mathrm{meas}}(\Omega)
    =
    S_{\mathrm{imp}}(\Omega)
    +
    S_{\mathrm{qba}}(\Omega)
    +
    S_{\mathrm{spin}}(\Omega)
    +
    S_{\mathrm{corr}}(\Omega),
\end{equation}
where the correlation term $S_{\mathrm{corr}}(\Omega)$ may vanish depending on the measurement configurations.
The corresponding noise-equivalent magnetic-field sensitivity is obtained by dividing the remaining noise $S_{\mathrm{meas}}(\Omega)$ by the field-to-output transduction factor,
\begin{equation}
    \delta B(\Omega)
    =
    \frac{\sqrt{S_{\mathrm{meas}}(\Omega)}}{\sqrt{\Gamma_S A_B}\,\abs{\rho_S(\Omega)}},
\end{equation}
with configuration-dependent modifications according to the quadrature rotation, the conditional suppression, and optical loss.
\section{Variational readout with vacuum probe}
We first consider  a variational readout with a vacuum probe. In this configuration, the detected optical quadrature is rotated away from the probe phase quadrature $\theta_{\mathrm{mag}} \neq 0$, which mixes probe amplitude fluctuations and backaction noise, thereby generating ponderomotive squeezing. 
The detected optical quadrature is
\begin{equation}
    Q^{\mathrm{out}}_{L} = P^{\mathrm{out}}_{L}\cos(\theta_{\mathrm{mag}})+X^{\mathrm{out}}_{L}\sin(\theta_{\mathrm{mag}}),
\end{equation}
Using the common noise decomposition introduced above, the variational-readout spectrum, including the output losses $\eta^{\mathrm{out}}_{\mathrm{mag}}$, can be written as
\begin{align}
S^{\mathrm{vr}}(\Omega) = &\eta^{\mathrm{out}}_{\mathrm{mag}}\left[S^{\mathrm{vr}}_{\mathrm{imp}}(\Omega)+S^{\mathrm{vr}}_{\mathrm{qba}}(\Omega) + S^{\mathrm{vr}}_{\mathrm{corr}}(\Omega)+S^{\mathrm{vr}}_{\mathrm{sig}}(\Omega)+S^{\mathrm{vr}}_{\mathrm{spin}}(\Omega)\right]+(1-\eta^{\mathrm{out}}_{\mathrm{mag}})S_{\mathrm{vac}}(\Omega),\\
    S^{\mathrm{vr}}_{\mathrm{imp}}(\Omega) = &S_{P_{L}^{\mathrm{in}}, P_{L}^{\mathrm{in}}}(\Omega)\cos^{2}(\theta_{\mathrm{mag}}) + S_{X_{L}^{\mathrm{in}}, X_{L}^{\mathrm{in}}}(\Omega)\sin^{2}(\theta_{\mathrm{mag}}),\\
    S^{\mathrm{vr}}_{\mathrm{qba}}(\Omega) = &\Gamma_{S}^{2}\abs{\chi_{S}(\Omega)}^{2} S_{X_{L}^{\mathrm{in}}, X_{L}^{\mathrm{in}}}(\Omega)\cos^{2}(\theta_{\mathrm{mag}}),\\
    S^{\mathrm{vr}}_{\mathrm{corr}}(\Omega) =&\Gamma_{S}\mathrm{Re}\left(\chi_{S}(\Omega) \right)S_{X_{L}^{\mathrm{in}}, X_{L}^{\mathrm{in}}}(\Omega)\sin(2\theta_{\mathrm{mag}}),\\
    S^{\mathrm{vr}}_{\mathrm{sig}}(\Omega) =&\Gamma_{S}A_{B}\abs{\rho_{S}(\Omega)}^2S_{B_{\mathrm{RF}}}(\Omega)\cos^{2}(\theta_{\mathrm{mag}}),\\
    S^{\mathrm{vr}}_{\mathrm{spin}}(\Omega)=&\left(\Gamma_{S}\gamma_{S}(\abs{\rho_{S}(\Omega)}^{2}+\abs{\chi_{S}(\Omega)}^{2}) (n_{S}+1/2) + \Gamma_{bb}\gamma_{bb}(\abs{\rho_{bb}(\Omega)}^{2}+\abs{\chi_{bb}(\Omega)}^{2}) (n_{bb}+1/2)\right)\cos^{2}(\theta_{\mathrm{mag}}).
\end{align}
 For the vacuum limited probing, only the losses after light-atom interaction degrade the ponderomotive squeezing and the measured magnetic signals. The variational-readout-enhanced magnetic sensitivity is therefore 
\begin{equation}
    \delta B^{\mathrm{vr}}_{y}(\Omega) = \frac{\sqrt{S^{\mathrm{vr}}_{\mathrm{imp}}(\Omega)+\eta^{\mathrm{out}}_{\mathrm{mag}}\left(S^{\mathrm{vr}}_{\mathrm{qba}}(\Omega)+ S^{\mathrm{vr}}_{\mathrm{corr}}(\Omega)+S^{\mathrm{vr}}_{\mathrm{spin}}(\Omega)\right)}}{\sqrt{\eta^{\mathrm{out}}_{\mathrm{mag}}\Gamma_{S}A_{B}}\abs{\rho_{S}(\Omega)}\cos(\theta_{\mathrm{mag}})} .
\end{equation}
This expression shows that variational readout improves sensitivity over a finite frequency range when the reduction in measurement-added noise exceeds the accompanying signal reduction from the quadrature rotation and detection loss. The achievable enhancement depends on the choice of readout angle $\theta_{\mathrm{mag}}$. In practice, for each analysis frequency, the sensitivity can be optimized using the frequency-dependent phase $\theta_{\mathrm{mag}}(\Omega)$ that minimizes the magnetic-field sensitivity across the sideband frequency.

\section{EPR-correlated probe and conditional quantum suppression}
We next consider a conditional quantum noise suppression approach using an EPR-entangled probe. In this configuration, the quantum fluctuations of the probe field interacting with the atomic ensemble are entangled with an auxiliary reference mode, enabling non-local correlations between the atomic measurement record and a separate optical channel.
For a two-mode squeezed input state, the quadrature fluctuations of each individual mode are amplified according to
\begin{align}
    S_{X_{L, \mathrm{mag}}^{\mathrm{in}},X_{L, \mathrm{mag}}^{\mathrm{in}}}(\Omega) = S_{P_{L,\mathrm{mag}}^{\mathrm{in}},P_{L,\mathrm{mag}}^{\mathrm{in}}}(\Omega)= S_{X_{L, \mathrm{EPR}}^{\mathrm{in}},X_{L, \mathrm{EPR}}^{\mathrm{in}}}(\Omega) = S_{P_{L,\mathrm{EPR}}^{\mathrm{in}},P_{L,\mathrm{EPR}}^{\mathrm{in}}}(\Omega) = \frac{\cosh(2r)}{2}, 
\end{align}
while the cross-channel correlations scale as $\sinh(2r)/2$. As a result, both imprecision and quantum backaction noise contributions increase by a factor $\cosh(2r)/2$, while the cross-correlations between the atomic and reference channels enable conditional noise reduction.
The atomic output is detected in the phase quadrature $P_{\mathrm{mag}}^{\mathrm{out}}$ ($\theta_{\mathrm{mag}} =0$), while the reference EPR field is measured in a rotated quadrature
\begin{equation}
    Q_{L,\mathrm{EPR}}^{\mathrm{out}}(\Omega) = P_{L,\mathrm{EPR}}^{\mathrm{out}}(\Omega)\cos(\phi_{\mathrm{EPR}})+X_{L,\mathrm{EPR}}^{\mathrm{out}}(\Omega)\sin(\phi_{\mathrm{EPR}}).
\end{equation}
The atomic output is linearly combined with the reference measurement,
\begin{equation}
    P_{L}^{\mathrm{cond}}(\Omega) = P_{L,\mathrm{mag}}^{\mathrm{out}}(\Omega) + g(\Omega)Q_{L,\mathrm{EPR}}^{\mathrm{out}}(\Omega),
\end{equation}
where $g(\Omega)$ is a frequency dependent gain.  
The combined output yields the conditional noise spectrum
\begin{equation}
    S^{\mathrm{cond}}(\Omega) = S_{\mathrm{mag}}(\Omega) + \abs{g(\Omega)}^{2}S_{\mathrm{EPR}}(\Omega) +2\mathrm{Re}\left(g(\Omega)S_{P_{L,\mathrm{mag}},Q_{L,\mathrm{EPR}}}(\Omega) \right).
\end{equation}
The atomic output spectrum used to fit the experimental data is given by (excluding the contribution of the magnetic signal $S_{\mathrm{sig}}(\Omega)$)
\begin{equation}
\begin{aligned}
S_{\mathrm{mag}}(\Omega)
&= \frac{1-\eta^{\mathrm{out}}_{\mathrm{mag}}}{2}
+\eta^{\mathrm{out}}_{\mathrm{mag}}
\Bigg[
\eta^{\mathrm{in}}_{\mathrm{mag}}\frac{\cosh(2r)}{2}
\left(1+\Gamma_S^2 \left|\chi_S(\Omega)\right|^2 \right)
\\
&\qquad\qquad
+\frac{1-\eta^{\mathrm{in}}_{\mathrm{mag}}}{2}
\left(1+\Gamma_S^2 \left|\chi_S(\Omega)\right|^2 \right)
+S_{\mathrm{spin}}(\Omega)
\Bigg],
\end{aligned}
\end{equation}
while the EPR reference channel spectrum is
\begin{equation}
     S_{\mathrm{EPR}}(\Omega)= \eta_{\mathrm{EPR}}\frac{\cosh(2r)}{2}+\frac{1-\eta_{\mathrm{EPR}}}{2},
\end{equation}
the complex cross-spectral density between the atomic and EPR reference channels is
\begin{equation}
      S_{P_{L,\mathrm{mag}}, Q_{L,\mathrm{EPR}}}(\Omega) = \sqrt{\eta^{\mathrm{out}}_{\mathrm{mag}}\eta^{\mathrm{in}}_{\mathrm{mag}}\eta_{\mathrm{EPR}}}\left[\cos(\phi_{\mathrm{EPR}} )+\Gamma_{S}\chi_{S}(\Omega)\sin(\phi_{\mathrm{EPR}})\right]\frac{\sinh(2r)}{2}.
\end{equation}
Minimizing the conditional spectrum with respect to the filter gain gives the optimal Wiener filter \cite{brown1997introduction,gould2021optimal,novikov2025hybrid},
\begin{equation}\label{eq-SM:g-opt}
   g_{\mathrm{opt}}(\Omega)=-\frac{S^{*}_{P_{L,\mathrm{mag}}, Q_{L,\mathrm{EPR}}}(\Omega)}{S_{\mathrm{EPR}}(\Omega)};
\end{equation}
we employ the non-causal Wiener filter, which estimates the signal from the entire data record (both past and future) and therefore admits the closed-form frequency-domain expression above; this is appropriate for the offline processing of the stationary data considered here. 
Using the expression~\eqref{eq-SM:g-opt}, the conditioned noise spectrum is rewritten as
\begin{equation}
      S^{\mathrm{cond}}(\Omega) = S_{\mathrm{mag}}(\Omega)-\frac{\abs{S_{P_{L,\mathrm{mag}}, Q_{L,\mathrm{EPR}}}(\Omega)}^2}{S_{\mathrm{EPR}}(\Omega)}.
\end{equation} 
The corresponding conditional noise-equivalent magnetic-field sensitivity is therefore
\begin{equation}
    \delta B^{\mathrm{cond}}_{y} = \frac{1}{\sqrt{\eta^{\mathrm{out}}_{\mathrm{mag}}\Gamma_{S}A_{B}}\abs{\rho_{S}(\Omega)}}\left[ \sqrt{S_{\mathrm{mag}}(\Omega)-\frac{\abs{S_{P_{\mathrm{mag}}, Q_{\mathrm{EPR}}}(\Omega)}^2}{S_{\mathrm{EPR}}(\Omega)}}\right].
\end{equation}
In contrast to variational readout, where correlations are generated within the measurement process, this configuration shows that quantum enhancement can also be achieved by subtracting measurement backaction and/or imprecision noise using non-local correlations with the reference beam.  The efficiency of this conditional noise reduction is determined by the strength of the cross-correlation quantified by the squeezing factor $r$.  
\section{Hybrid configuration combining variational readout and EPR conditioning}
The hybrid configuration combines a rotated atomic detection quadrature with conditional noise subtraction based on the EPR-correlated reference channel. 
The measured atomic and EPR channel quadratures are
\begin{align}
    Q^{\mathrm{out}}_{L, \mathrm{mag}}(\Omega) &= P^{\mathrm{out}}_{L, \mathrm{mag}}(\Omega)\cos(\theta_{\mathrm{mag}})+X^{\mathrm{out}}_{L, \ \mathrm{mag}}(\Omega)\sin(\theta_{\mathrm{mag}}),\\
    Q_{L,\mathrm{EPR}}^{\mathrm{out}}(\Omega) &= P_{L,\mathrm{EPR}}^{\mathrm{out}}(\Omega)\cos(\phi_{\mathrm{EPR}})+X_{L,\mathrm{EPR}}^{\mathrm{out}}(\Omega)\sin(\phi_{\mathrm{EPR}}).
\end{align}
The hybrid conditional spectrum retains a similar conditional structure,
\begin{equation}
    S^{\mathrm{hyb}}(\Omega)
    =
    S_{\mathrm{mag}}^{\mathrm{vr}}(\Omega)
    +
    |g_{\mathrm{opt}}^{\mathrm{hyb}}(\Omega)|^2 S_{\mathrm{EPR}}(\Omega)
    +
    2\mathrm{Re}\!\left[
    g_{\mathrm{opt}}^{\mathrm{hyb}}(\Omega)
    S_{Q_{L,\mathrm{mag}},Q_{L,\mathrm{EPR}}}^{\mathrm{hyb}}(\Omega)
    \right],
\end{equation}
where $S_{\mathrm{mag}}^{\mathrm{vr}}(\Omega)$ is the variational-readout atomic spectrum introduced in Sec. V, but here the noise is amplified by $\frac{\cosh(2r)}{2}$ [excluding the contribution of the magnetic signal $S_{\mathrm{sig}}(\Omega)\cos^{2}(\theta_{\mathrm{mag}})$]
\begin{equation}
    \begin{aligned}
        S^{\mathrm{vr}}_{\mathrm{mag}}(\Omega)
   &= \frac{1-\eta^{\mathrm{out}}_{\mathrm{mag}}}{2}
   + \eta^{\mathrm{out}}_{\mathrm{mag}} \Big[
      \eta^{\mathrm{in}}_{\mathrm{mag}}\frac{\cosh(2r)}{2}\big(1+\Gamma_{S}^{2}\abs{\chi_{S}(\Omega)}^{2}\cos^{2}(\theta_{\mathrm{mag}})+\Gamma_{S}\mathrm{Re}\left(\chi_{S}(\Omega) \right)\sin(2\theta_{\mathrm{mag}})\big) \\
   &\quad + \frac{1 - \eta^{\mathrm{in}}_{\mathrm{mag}}}{2}
      \big(1 + \Gamma_{S}^{2}\abs{\chi_{S}(\Omega)}^{2}\cos^{2}(\theta_{\mathrm{mag}})
      + \Gamma_{S}\mathrm{Re}\left(\chi_{S}(\Omega) \right)\sin(2\theta_{\mathrm{mag}})\big) + S_{\mathrm{spin}}(\Omega)\cos^{2}(\theta_{\mathrm{mag}})
   \Big].  
    \end{aligned}
\end{equation}
The noise spectrum for the EPR conditioning reference remains the same, while the hybrid cross-correlation is
\begin{equation}
    S^{\mathrm{hyb}}_{Q_{L,\mathrm{mag}}, Q_{L,\mathrm{EPR}}}(\Omega)
   = \sqrt{\eta^{\mathrm{out}}_{\mathrm{mag}} \eta^{\mathrm{in}}_{\mathrm{mag}} \eta_{\mathrm{EPR}}}
   \left[
      -\cos(\phi_{\mathrm{EPR}}+\theta_{\mathrm{mag}})
      + \Gamma_{S}\chi_{S}\sin(\phi_{\mathrm{EPR}})\cos(\theta_{\mathrm{mag}})
   \right]\frac{\sinh(2r)}{2}.
\end{equation}
and adapted to the optimal Wiener gain
\begin{equation}
    g^{\mathrm{hyb}}_{\mathrm{opt}}(\Omega) = -\frac{S^{\mathrm{hyb},*}_{Q_{L,\mathrm{mag}}, Q_{L,\mathrm{EPR}}}(\Omega)}{S_{\mathrm{EPR}}(\Omega)}.
\end{equation}
The resulting sensitivity for the hybrid configurations is expressed as 
\begin{equation}
    \delta B^{\mathrm{hyb}}_{y} = \frac{1}{\sqrt{\eta^{\mathrm{out}}_{\mathrm{mag}}\Gamma_{S}A_{B}}\abs{\rho_{S}(\Omega)}\cos(\theta_{\mathrm{mag}})}\left[ \sqrt{S^{\mathrm{vr}}_{\mathrm{mag}}(\Omega)-\frac{\abs{S_{Q_{L,\mathrm{mag}}, Q_{L,\mathrm{EPR}}}(\Omega)}^2}{S_{\mathrm{EPR}}(\Omega)}}\right].
\end{equation}

The hybrid scheme combines intra-channel correlations generated by variational readout with cross-channel correlations provided by the EPR reference. This expands the accessible correlation space of the measurement, enabling improved noise cancelation beyond either approach alone.

\FloatBarrier
\section{Optimal magnetic-field sensitivity across measurement configurations}
To assess the best achievable performance, we compare the magnetic-field sensitivity obtained from different measurement configurations, including variational readout, EPR conditioning, and their hybrid combination, using calibrated system parameters.
The optimal sensitivity is defined as the minimum value at each sideband frequency across the corresponding sensitivity curves (indicated by different markers), corresponding to frequency-dependent optimization of the atomic detection quadrature and the EPR conditioning phase.
The resulting solid curves represent the envelope of the accessible sensitivity spectra and serve as a benchmark for the maximum achievable quantum enhancement under the given experimental conditions. The results are summarized in Fig.~S3.
\begin{figure*}[htp]
\centering
\includegraphics[width=0.8\textwidth]{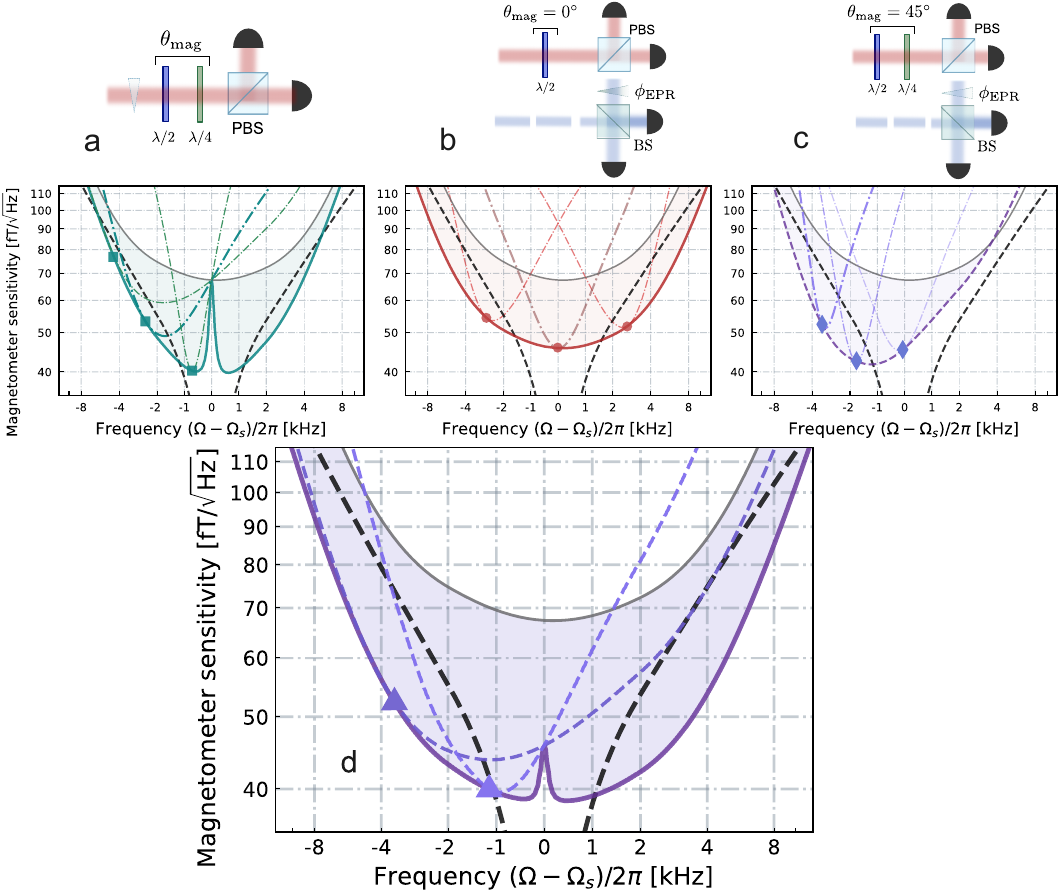}
\caption{\textbf{Broadband enhancement of a magnetometer sensitivity via quantum-noise engineering.}  The solid gray curve denotes the reference sensitivity of a standard orientation magnetometer. Regions below this reference (highlighted by coloured areas) indicate quantum-enhanced performance. \textbf{a}, Sensitivity obtained with variational readout for three fixed atomic detection angles, ($\theta_{\mathrm{mag}}: -30^{\circ}, -60^{\circ}, -80^{\circ}$) (dash-dotted curves). The solid teal curve shows the optimal sensitivity obtained by frequency-dependent tuning of the atomic detection angle. \textbf{b}, Sensitivity obtained with EPR conditioning for three entangled-channel detection phases ($\phi_{\mathrm{EPR}}: -30^{\circ}, 0^{\circ}, 30^{\circ}$), with the atomic channel fixed at the phase quadrature ($\theta_{\mathrm{mag}}: 0^{\circ}$). The solid red curve shows the optimal sensitivity obtained by tuning the EPR phase. \textbf{c}, Sensitivity in the hybrid configuration, combining a fixed  variational readout ($\theta_{\mathrm{mag}}: -55^{\circ}$) with three EPR conditioning phases ($\phi_{\mathrm{EPR}}: 0^{\circ}, 90^{\circ}, 130^{\circ}$). The dashed purple curve shows the sensitivity for this fixed variational readout angle with optimized EPR conditioning. \textbf{d}, Hybrid configurations evaluated for two fixed variational readout angles, illustrating how quantum-noise engineering reshapes the sensitivity across different frequencies. The solid purple curve indicates the optimal hybrid sensitivity obtained by frequency-dependent optimization of both the atomic and EPR detection phases. Markers (square, circle, diamond, triangle) indicate the frequencies at which each configuration reaches its minimum sensitivity. The dashed black curve represents the standard quantum limit. The optimal solid curves are obtained by selecting the minimum sensitivity at each frequency across the explored parameter range. The atomic Larmor frequency is set to $\Omega_{S}/2\pi= 100$\,kHz.} 
\label{fig:quantum enhanced engineering approaches}
\end{figure*}
\section{Comparison between single-mode and EPR injection.}
Parallel EPR conditioning offers several advantages over single-mode squeezed-vacuum injection. Although it comes with a 3\,dB penalty for the same squeezing factor, it is less sensitive to optical loss and less affected by anti-squeezing noise with phase jitter \cite{Danilishin2019}. The presence of an auxiliary correlated reference channel enables frequency-dependent complex gain and optimal Wiener filtering \cite{gould2021optimal, novikov2025hybrid}, allowing partial compensation of imperfect frequency response. These features make the EPR scheme particularly advantageous in broadband and non-ideal regimes, and such imperfections may be mitigated using continuous-variable entanglement purification protocols based on QND measurements \cite{duan2000entanglement}.

\FloatBarrier
\section{Extending the Spin susceptibility function with Voigt correction}
In the data analysis, we observe small discrepancies between the experimental spectra and the theoretical fits within $\pm 500$\,Hz of the Larmor frequency. Such discrepancies can be reduced by replacing the atomic susceptibility with a Voigt profile \cite{voigt}. The Voigt model provides a possible explanation in terms of additional inhomogeneous broadening beyond the intrinsic Lorentzian response.
This correction does not affect the main conclusions of the work, but improves the agreement between the model and experimental data in the near-resonant region as shown in Fig.~S4.
\begin{figure*}[htp!]
    \centering
    \includegraphics[width=0.4\linewidth]{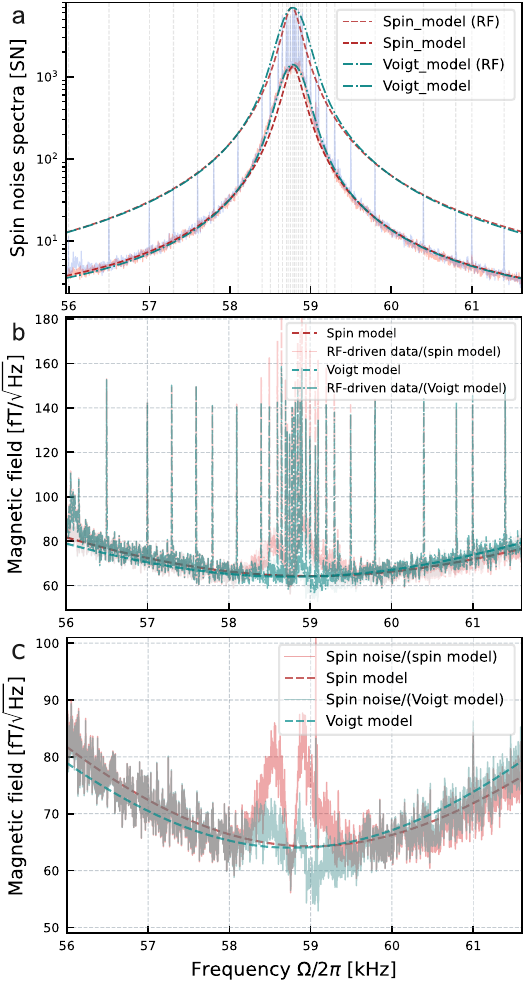}
    \caption{\textbf{Spin transfer function modelling with Lorentzian and Voigt susceptibilities.} \textbf{a}, The spin-noise spectrum and the multi-tone RF-driven response are fitted using a Lorentzian model (red dashed curve) and a Voigt profile (teal curve). The Voigt model, which convolves the Lorentzian with an additional Gaussian component, provides a better description of the measured spin-noise lineshape. In this Voigt model, the Lorentzian and Gaussian contributions are assumed to have the same characteristic width. \textbf{b \& c}, When the spectra are converted into noise-equivalent magnetic amplitude spectral density, the discrepancy between the theoretical model and experimental data is significantly reduced when using the Voigt profile. }
    \label{fig:enter-label}
\end{figure*}
\FloatBarrier
\bibliography{references}